\documentclass[12pt,a4paper]{article}
\pdfoutput=1
\usepackage[latin1]{inputenc}
\usepackage{amsfonts,amsbsy,bm,euscript,mathrsfs}
\usepackage{amssymb,stmaryrd,faktor,slashed}
\usepackage{color}
\usepackage[tbtags]{amsmath}
\usepackage[bookmarks=true,colorlinks=true,linkcolor=black,citecolor=black,urlcolor=black,bookmarksnumbered]{hyperref}
\usepackage[nodayofweek]{date time}

\usepackage[a4paper,text={170mm,257mm},centering]{geometry}

 \usepackage{authblk, physics, bbold, multirow, comment}

\usepackage{graphicx}
\usepackage[verbose]{wrapfig}
\usepackage{caption}

\numberwithin{equation}{section}

\makeatletter
\renewcommand\section{\@startsection {section}{1}{\z@}%
	{-3.5ex \@plus -1ex \@minus -.2ex}%
	{2.3ex \@plus.2ex}%
	{\normalfont\large\bfseries}}
\renewcommand\subsection{\@startsection{subsection}{2}{\z@}%
	{-3.25ex\@plus -1ex \@minus -.2ex}%
	{1.5ex \@plus .2ex}%
	{\normalfont\normalsize\bfseries}}
\makeatother

\expandafter\def\expandafter\bfseries\expandafter{\bfseries\ifmmode\else\boldmath\fi}
\expandafter\def\expandafter\mdseries\expandafter{\mdseries\ifmmode\else\unboldmath\fi}
\expandafter\def\expandafter\normalfont\expandafter{\normalfont\ifmmode\else\unboldmath\fi}

\providecommand{\href}[2]{#2}

\newcommand{\mathsym}[1]{{}}
\def\id{\protect{{1 \kern-.28em{\rm l}}}}
\def\be{\begin{equation}}
\def\ee{\end{equation}}

\def\ha{\tfrac{1}{2}}

\def\ci{\cite}
\def\N{{\mathcal N}}

\def\S{{\mathcal S} }

\def\a{\alpha}

\def\g{\gamma}

\def\k{\kappa}
\def\four{\tfrac14}

\def\Tr{{\rm Tr}}

\def\l {\lambda}

\def\m{\mu}

\def\foot{\footnote}
\newcommand{\rf}[1]{(\ref{#1})}

\def\F{{\cal F}}

\def\no{\nonumber}
\def\J{{\cal J}}

\def\la{\label}
\def\l{\lambda}

\def\adss{$AdS_5 \times S^5$\ }

\def\r{\rho}

\def\varpi{{\rm w}}

\def\Z{{\cal Z}}

\def\del{\partial}
\def\s{\sigma}
\def\eps{{\epsilon}}
\def\n{\nu}

\def\edd{\end{document}}

\def\iffa{\iffalse}

\def\te{\textstyle}

\def\dd {{\rm d}}

\def\sql{{\sqrt{\l}}}

\def\qq{{\rm q}}
\def\rk {{\rm u}}

\def \ov {\over}

\def \k {\varkappa} 
\def \dt {\dt}

\def \kk {{\rm k}}

\begin{document}

	\begin{flushright}\small{Imperial-TP-AT-2021-{05}}
	\end{flushright}
	\vspace{2.0cm}
	\begin{center}
		
		{\Large\bf On type 0 string theory in solvable RR backgrounds
			\vspace{0.3cm}
		}
		
		\vspace{1.0cm}
		
		{Torben Skrzypek$^{a,}$\footnote{t.skrzypek20@imperial.ac.uk} and
			Arkady A. Tseytlin$^{a,}$\footnote{Also at the Institute of Theoretical and Mathematical Physics (ITMP) of Moscow U. and
				Lebedev Institute.
				
				\ \ tseytlin@imperial.ac.uk}
		}

		\vspace{0.5cm}	
		
		{\em
			$^{a}$Blackett Laboratory, Imperial College, London SW7 2AZ, U.K.
		}
		
	\end{center}

	\vspace{0.5cm}
	
\begin{abstract}
Motivated by a possibility of solving non-supersymmetric type 0 string theory 
in $AdS_5 \times S^5$ background using integrability, 
we revisit the construction of type 0 string spectrum 
in some solvable examples of backgrounds with RR fluxes
that are common to type IIB and type 0B theories. 
The presence of RR fluxes requires the use of a 
Green-Schwarz description for type 0 string theory. 
Like in flat space, the spectrum of type 0 theory 
can be derived from the type II theory spectrum 
by a $(-1)^F$ orbifolding, i.e. combining the untwisted sector where GS fermions are periodic with the 
twisted sector where GS fermions are antiperiodic (and projecting out all spacetime fermionic states). 
This construction of the 
type 0 spectrum may also be implemented using a Melvin background that allows to continuously 
interpolate between the type II and type 0 theories. As an illustration, we discuss the type 0B spectrum in the pp-wave background 
which is the Penrose limit of \adss with RR 5-form flux and also in the pp-wave background 
which is the Penrose limit of $AdS_3 \times S^3 \times{ \rm T}^4$ supported by mixed RR and NSNS 3-form fluxes. 
We show that increasing the strength of the RR flux increases the value of the effective 
normal ordering constant (which 
determines the mass of the type 0 tachyon in flat space) 
and thus effectively decreases the 
 momentum-space domain 
of instability of the ground state. 
We also comment on the semiclassical sector of states of type 0B theory in $AdS_5 \times S^5$. 
		\end{abstract}

	\newpage
	\tableofcontents
	
\def \Z {{\cal Z}}
	\def \OO {{\cal O}}
	\def \kk { k}
	\def \bq {{q'}} \def \barp {x} \def \barq {y}
	\def \rk {{\rm k}}
	\setcounter{footnote}{0}
	\setcounter{section}{0}
	\def \ed { \small \bibliographystyle{JHEP} \bibliography{Type.bib} \end{document}}
\def \II {{\cal I}}
\def \rX {{\rm X}}
 \def \S {{\cal S}}
 \def \mm {{\rm m}}
\def \AA {{\cal A}}
\def \bA {{\bar A}}
\def \xx {{\rm x}} \def \rT {{\rm T}}
\def \tZ {{\widetilde \Z}}
\def \k {\kappa} \def \rk {{\rm w}}
\def \PP {{\rm P}}
\def \qq {{\rm q}} 
\def \mub {\hat \mu}
\def \na {\nabla}

\

\section{Introduction}\label{intro}
It was suggested some time ago \ci{Polyakov:1998ju,Klebanov:1998yya,Klebanov:1999ch} that type 0 string theory 
\ci{Dixon:1986iz,Seiberg:1986by,Thorn} may provide an example of planar AdS/CFT duality in a non-supersymmetric setup. 
The closed 
type 0B theory admits the same classical \adss solution with RR 5-form flux 
 as the type IIB theory but a major problem is the presence of the tachyon 
 $T$ (with flat-space mass $m^2_0 = - {2\ov \a'}$). 
 The coupling of $T$ to the RR field strength \ci{Klebanov:1998yya}
 implies that 
 the mass of the tachyon may be shifted by the presence of RR flux and thus the 
 tachyon may disappear for some 
 small enough critical value of the 
 effective string tension $\sqrt \l \equiv {R^2_{\rm AdS} \ov \a'}$ \ci{Klebanov:1999ch,Klebanov:1999um}.\foot{From the weak-coupling gauge theory side the presence of the closed string tachyon 
 may be seen as the appearance of an imaginary part of the anomalous dimension of the dual operator at some large enough critical value of the 't Hooft coupling $\l$ 
 \ci{Klebanov:1999um}. 
 The suggestion of duality \ci{Klebanov:1999ch} between type 0 string theory 
 in \adss and non-supersymmetric gauge theory on self-dual 
 type 0 D3 branes 
 \ci{Bergman:1997rf} (which is a $(-1)^F$-type orbifold of $\N=4$ SYM \ci{Klebanov:1999ch,Nekrasov:1999mn})
 has of course several known caveats (even in the strict planar limit).
 One may view this type 0 example as being analogous to non-supersymmetric 
 $AdS_5 \times S^5/\Gamma$ orbifolds of type IIB theory \ci{Kachru:1998ys}.
 In all of these cases there is also a problem of generating 
 new double-trace terms in the action with couplings having complex values at zeroes of the corresponding beta-functions 
 \ci{Tseytlin:1999ii,Adams:2001jb,Armoni:2003va,Dymarsky:2005nc,Pomoni:2008de}. These 
 may be interpreted in terms of tachyons appearing at small enough gauge coupling
 or small enough effective string tension. 
 Here we shall focus only on the type 0 closed string tachyon present at large string tension 
 and the dependence of its mass on the strength of the RR background. }

 Testing 
 this conjecture requires an 
 understanding of the spectrum of type 0 string theory in \adss background beyond the large 
 tension $\sql \gg 1$ limit. In the case of type IIB string theory 
 in \adss at genus zero 
 (dual to planar $\N=4$ SYM) 
 the spectrum of string states is at least in principle computable using integrability (see, e.g., \ci{Beisert:2010jr,vanTongeren:2013gva,Gromov:2014caa,Gromov:2017blm}). 
 Since the bosonic part of the \adss string worldsheet 
 action (shared by type 0 and type II theories) is integrable, 
 one may conjecture that integrability may also apply 
 to the type 0 case.
 Let us note that 
 the spectral problem for integrable superstrings on (non-supersymmetric) orbifolds and TsT transformations of \adss 
 was addressed in \ci{deLeeuw:2012hp,vanTongeren:2013gva,Kazakov:2015efa,Levkovich-Maslyuk:2020rlp} but the case of type 0 theory 
 remains to be treated explicitly. 
 
 As \adss is supported by the RR flux, 
 this requires formulating the type 0 string action in the Green-Schwarz (GS) 
 approach. 
 In flat space the type 0 string has a straightforward description 
 in the NSR formalism 
 (involving a diagonal GSO projection which excludes spacetime fermions) \ci{Dixon:1986iz,Seiberg:1986by,Polchinski:1998rr}. 
 Using 
 light-cone gauge it can also be described in the GS formalism as a
 $(-1)^{F}$ orbifold of type II string theory, 
 i.e. as a combination of the two sectors -- with periodic
 and antiperiodic GS fermions \ci{Dixon:1986iz}.
 How to generalise this construction 
 to the case of non-trivial RR backgrounds like \adss 
 without applying any compactification and limits may first appear to
 be non-trivial.

 Motivated by the example of the Melvin background that allows to interpolate between the 
 type II and type 0 theories \ci{Russo:1995ik,Tseytlin:2001qb,Takayanagi:2001jj}
 we shall argue that starting with the type II superstring in a consistent RR background 
 one can find the spectrum of the 
 type 0 string in this 
 background by the same orbifolding prescription as in flat space -- as a 
 combination of the untwisted sector 
 (described by the type II GS string action with periodic worldsheet fermions but all spacetime fermions projected out from the spectrum) 
 and the twisted sector (described by the 
 type II GS action with antiperiodic worldsheet fermions). 
 
 Our aim below 
 will be to discuss a 
 few simple examples of solving 
 type 0 string theory 
 in pp-wave backgrounds with RR fluxes 
 using the 
 GS formulation where the 
 connection to the corresponding solution of type II theory is straightforward.
 We shall demonstrate how the orbifolding construction works and 
 draw some lessons that may be useful in the \adss case.
 From the exact expression of the 
 type 0 string spectrum available in these cases we shall see 
 that the presence of the RR fluxes indeed increases the value of the effective tachyon mass-squared 
 compared to its negative flat-space value. 
 
 \
 
 We shall start in section 2 with reviewing the solution of type II string theory 
 in Melvin background and explain the relation to 
 the corresponding spectrum of type 0 string theory. 
 
 In section 3 we shall use the known solution \ci{Metsaev:2001bj,Berenstein:2002jq,Metsaev:2002re} 
 of type IIB string theory 
 in the pp-wave background which is the Penrose limit of 
 \adss 
 and the above orbifolding prescription to construct the corresponding type 0 string 
 spectrum \ci{Bigazzi:2002gw} and the modular-invariant partition function 
 \ci{Takayanagi:2002pi}. 
 We shall find the explicit dependence of the tachyon mass on the strength of the RR field and show that 
 it approaches zero in the limit of infinite flux. 
 
 In section 4 we shall perform a similar analysis in the case of the pp-wave background which is the Penrose limit of
 $AdS_3 \times S^3 \times {\rm T}^4$ supported by a combination of RR and NSNS 3-form fluxes 
 (with the solution of the corresponding type IIB theory found in \ci{Berenstein:2002jq,Russo:2002rq}).
 Here 
 we will see again that the type 0 tachyon mass-squared is shifted up by the RR flux. 
 
 Some comments on 1-loop corrections to the energy of 
 semiclassical states in the type 0B spectrum in \adss will be made in section 5.

 In Appendix \ref{A} we shall review the construction of the 
 type 0 string spectrum in flat space. 
 In Appendix \ref{B} we shall consider type II string theory 
 in pp-wave background with an extra Melvin twist in a plane
 and recover from its solution the pp-wave type 0 string spectrum by taking a special limit.\foot{
 One may also consider starting directly with type II theory in \adss (rather than its pp-wave limit) 
 and add a Melvin twist (cf. \ci{Ganor:2007qh,Dhokarh:2008ki}
 for related constructions) in order to interpolate to type 0 theory.
 However, solving such an interpolating string theory (even if it may still be integrable) will be non-trivial.
 }
 In Appendix \ref{C} we shall demonstrate how to reproduce the lower-level part of the 
 type 0 spectrum in pp-wave background discussed in section 3 by 
 expanding the type 0B low-energy effective action to quadratic order in fluctuations.

\section{From type II to type 0 theory via Melvin twist} \label{Melvin}

The construction of type 0 string theory in flat space as a $(-1)^F$ orbifold of type II theory 
in GS description \cite{Dixon:1986iz} 
reviewed in Appendix A suggests that the same recipe may apply also in general curved backgrounds that are common to type II and type 0 theories. The full type 0 spectrum will be a sum of untwisted and twisted sectors. 
 The untwisted sector is obtained from the corresponding type II spectrum by projecting 
 out all spacetime fermion states. The twisted sector should be found by starting again with the type II GS action and now taking fermions to be antiperiodic on the cylinder.\foot{As 
 is well known, the GS action in generic type II background is $\kappa$-symmetric 
 if the background solves the corresponding string low-energy (or supergravity) equations. 
 The same should be true also in the type 0 case provided the GS string is coupled only to 
 the massless fields that are common to type II and type 0 theories
 (i.e. the background values of the type 0 tachyon field and the second copy of the RR fields should be set to zero).}
 
To further clarify this relation between the type II and type 0 theories, 
 here we shall discuss the type II solution in Melvin background 
 that allows to continuously interpolate between the type II and type 0 spectra \cite{Russo:1995ik,Takayanagi:2001jj}.\foot{Here
 we will consider only perturbative 10d string theory setting. 
Relations between type II and type 0 theories were also discussed from 11d M-theory perspective in 
\ci{Bergman:1999km,Costa:2000nw,Russo:2001tf}. In particular, ref. \ci{Russo:2001tf}
discussed the spectrum and the tachyon mass in the interpolating background.
}
The flux tube or KK Melvin background is represented by the locally flat 10d 
metric\foot{The Kaluza-Klein generalization of the Melvin solution was introduced in \ci{Gibbons:1986cq,Gibbons:1987ps}
and its interpretation as flat 5d space with a specific identification was given in \ci{Dowker:1993bt,Dowker:1994up}.
The embedding of the Melvin solution into string theory was done in \ci{Tseytlin:1994ei,Russo:1995tj,Russo:1995ik}.}
\begin{equation}\la{221}
\dd s^2= \dd s^2_{1,6} + \dd y^2 + \dd \rho^2 + \rho^2(\dd \varphi + q\, \dd y)^2 \ , 
\end{equation}
 where $y$ is compactified on a circle of radius $R$ and $q$ is an arbitrary parameter.
 Moving around the $y$-circle 
 induces a rotation of $2\pi R q$ in the $(\rho,\varphi)$ plane. When 
\begin{equation}\la{222}
\xi = qR
\end{equation}
 is integer valued, this rotation does not affect bosonic fields but changes the sign of the fermions and thus supersymmetry is broken 
 unless $\xi$ is even; in the latter case this background is topologically trivial and the resulting superstring spectrum is the same as in flat space.

 Fixing the light-cone gauge, 
the corresponding GS action may be written as \cite{Russo:1995ik}
\begin{equation}\la{223}
\S=\frac{1}{\pi\alpha'}\int\dd^2\sigma~ \left(G_{\mu\nu}\partial_+x^\mu\partial_-x^\nu + iS_R\mathcal{D}_+S_R+iS_L\mathcal{D}_-S_L\right)\ ,
\end{equation} 
where $\mathcal{D}_{\pm}=\partial_{\pm}+\frac{1}{4}\omega^{mn}_\mu \Gamma_{mn}\partial_{\pm}x^\mu$ with $\omega^{12}=-\omega^{21}=-q\dd y$.
Using $x=x_1+ix_2=\rho e^{i\varphi}$ and setting 
\begin{equation}\la{224}
x=e^{-iqy }X\ , \qquad\qquad S_{R,L}=e^{\mp \frac{iq}{2}y }\Lambda_{R,L}
\end{equation}
one finds that $X$ and $\Lambda_{R,L}$ satisfy free equations of motion. 
Since $y$ is compact, the string can wind $w$ times around it and then the prefactors in \rf{224} pick up a phase $\exp(-2\pi i \gamma )$ or $\exp(-\pi i \gamma)$ under a shift $\sigma\to\sigma+2\pi$, where 
\begin{equation}\la{225}
	\gamma=wqR=w\xi\,.
\end{equation}
Since the fields $x$ and $S_{R,L}$ should be periodic in $\s$, these phases are to be compensated 
 by twisting periodicities of $X$ and $\Lambda_{R,L}$. 
 This results in frequencies $(n\pm \gamma)$ for bosonic modes and $(n\pm \ha \gamma)$ for the fermionic
 ones.
 Our notation follow \cite{Russo:1995tj,Russo:1995ik}, i.e. we label the creation and annihilation operators $\alpha_{n\pm}^\dagger$, $\alpha_{n\pm}$ for bosons and $\beta_{n\pm}^\dagger$, $\beta_{n\pm}$ for fermions 
 (with canonical normalisation $[\alpha_{n-},\alpha_{n-}^\dagger]=\delta_{nm}$, etc.).
 The 
 mass operator may be written as \cite{Russo:1995ik}
\begin{equation}\la{226}
\alpha'M^2=\frac{\alpha'}{R^2}\hat \mm^2 +\frac{R^2}{\alpha'}w^2+2\big(\hat{N}_R+\hat{N}_L+A\big)\ , \qquad \qquad 
\hat \mm \equiv \mm-\xi(\hat{J}_b+\ha \hat{J}_f)\ , 
\end{equation}
where $\mm$ is the integer momentum in $y$ direction, 
which is shifted in $\hat \mm$ by the angular momenta of bosonic $\hat{J}_b$ and fermionic modes $\hat{J}_f$.
 $A$ is the normal ordering constant. Explicitly, we have 
\begin{align}\no
\hat{J}_b&= \tilde{\alpha}_0^\dagger\tilde{\alpha}_0-\alpha_0^\dagger \alpha_0 + \sum_{n=1}^\infty\te \big(\alpha_{n+}^\dagger \alpha_{n+} - \alpha_{n-}^\dagger \alpha_{n-}+\tilde{\alpha}_{n+}^\dagger \tilde{\alpha}_{n+}- \tilde{\alpha}_{n-}^\dagger \tilde{\alpha}_{n-}\big)\ , \\\la{228}
\hat{J}_f
&=\tilde{\beta}_{0}^\dagger\tilde{\beta}_{0}-\beta_{0}^\dagger \beta_{0 } + \sum_{n=1}^\infty\big(\beta_{n+}^\dagger \beta_{n+}-\beta_{n-}^\dagger \beta_{n-}+\tilde{\beta}_{n+}^\dagger \tilde{\beta}_{n+}-\tilde{\beta}_{n-}^\dagger\tilde{\beta}_{n-}\big)\ ,\\
\hat{N}_{R}&=\sum_{n=1}^{\infty}(n-\gamma)\alpha_{n+}^\dagger \alpha_{n+}+\sum_{n=0}^{\infty}(n+\gamma)\alpha_{n-}^\dagger \alpha_{n-}+\sum_{n=1}^\infty{\te \left(n-\frac{\gamma}{2}\right)}\beta_{n+}^\dagger \beta_{n+}+\sum_{n=0}^\infty{\te \left(n+\frac{\gamma}{2}\right)}\beta_{n-}^\dagger \beta_{n-}\ , \no \\ \no \la{230}
\hat{N}_{L}&=\sum_{n=1}^{\infty}(n-\gamma)\tilde{\alpha}_{n-}^\dagger \tilde{\alpha}_{n-}+\sum_{n=0}^{\infty}(n+\gamma)\tilde{\alpha}_{n+}^\dagger \tilde{\alpha}_{n+}+\sum_{n=1}^\infty{\te \left(n-\frac{\gamma}{2}\right)}\tilde{\beta}_{n-}^\dagger \tilde{\beta}_{n-}+\sum_{n=0}^\infty\te \left(n+\frac{\gamma}{2}\right)\tilde{\beta}_{n+}^\dagger\tilde{\beta}_{n+}\ .\no 
\end{align}
Here the tilde denotes left-moving modes and the spinor indices as well as 
the contributions of 6 flat bosonic 
dimensions have been suppressed. 
For $0 \leq \gamma\leq 1$, the normal ordering constant is given in terms of the Hurwitz zeta-function 
$\zeta(s,a)=\sum_{n=0}^\infty (n+a) ^{-s}$ as
\begin{equation}\la{231}
A(\gamma)=2\, \zeta(-1, \gamma)-8\, \zeta(-1, \ha \gamma)+6\, \zeta(-1,1)=-\gamma\ , \qquad \qquad \gamma\in[0,1] \ . 
\end{equation}
When $1 \leq \gamma < 2 $ the energy contribution of the bosonic mode of frequency $(1-\gamma)$ becomes negative and has to be reordered; this results in $A(\gamma)=- 2+\gamma$.
 At $\gamma=2$ both bosonic and fermionic modes are to be reshuffled and we can absorb the $\gamma$-shift in renaming of all modes, 
 getting back to the flat-space spectrum. This is exactly 
 what happens in the globally trivial case of $\xi=2$ for any value of $w$. 
 Extended to all values of $\g$, $A(\g) $ looks like a triangle wave, oscillating between values 0 and -1. 
 Thus for non-trivial cases of $\g\notin 2\mathbb{Z}$, 
 we find tachyonic ground states in the winding sectors and 
 broken supersymmetry \ci{Tseytlin:2001qb}. 

 Let us now consider the limits $R\to\infty$ and $R\to0$ while keeping $\xi$ fixed. 
 For $R\to\infty$ the KK momentum $p^y= \mm/R$ becomes continuous while the winding modes become infinitely heavy. 
 Thus we end up with the $w=0$ sector and recover the type II string theory on Minkowski space. 

Considering $R\to0$ and specifying to the $\xi=1$ case, 
 the periodicity in $\gamma =w \xi $ implies that we have to split the spectrum in two sectors, an ``untwisted" one 
 with even $\gamma$ and a ``twisted" one 
 with odd $\gamma$. Then taking the limit $R\to 0$, 
 the winding number $w$ becomes effectively continuous (i.e. is replaced by the momentum of the T-dual coordinate 
 $p^{\tilde y} = { w/ \tilde R}, \ \tilde R= {\a'/ R}$) 
 and we get 
 just two distinct sectors of states. 

Furthermore, in the limit $R\to0$ the first ``KK-momentum" 
term in \rf{226} blows up unless $\hat \mm = \mm-\xi(\hat{J}_b+\ha \hat{J}_f)$ 
 vanishes.
 As $\mm$ is integer, for $\xi=1$ we can find a solution of the condition $\hat \mm=0$ only if 
 $\hat{J}_f$ in a given state is even.
 This projects out all fermionic states from the spectrum, i.e. the only states that have finite masses in the $R\to0$ limit are the bosonic ones. 
 The untwisted sector thus consists of the bosonic spectrum of the T-dual type II theory. The 
 twisted sector has bosonic states with a tachyonic ground state ($\a' m_0^2 =2A=-2$) 
 and fermionic states corresponding to odd numbers of excitations of the half-integer frequency modes; 
imposing the 
 level-matching condition the latter are projected out and we end up with the bosonic states only (even before imposing the $\hat \mm =0$ constraint). 

 The resulting spectrum is exactly the same as that 
 of type 0 string theory in flat space. 
 Thus we can obtain the 
 type 0 spectrum from the 
 type II one via Melvin twist and T-duality.\foot{Note that this relation between type II and type 0 theories is not ``mutual": we cannot 
 obtain the type II spectrum from the 
 type 0 one by reversing this procedure.}
This construction can be used to find the type 0 spectrum from the 
type II one on non-trivial backgrounds that are solutions to both type II and type 0 theories
(assuming these backgrounds admit a translational as well as rotational isometries so that one can compactify one dimension and apply the 
Melvin twist in a 
2-plane).
 Then the above transformation at $\xi=1$ will result in the type 0 string theory on the T-dual background.
 Thus at least for the backgrounds allowing the introduction of a Melvin twist, the type 0 spectrum
 can be obtained from the type II one by the same orbifolding procedure as in flat space \cite{Dixon:1986iz}.

One immediate application of the above discussion 
 is to derive the type 0 spectrum on the Melvin background. The untwisted sector will be given by the bosonic part of the type II spectrum.
In the twisted sector the mass operator is the same as in \rf{226} but since 
we have to take the GS fermions to be antiperiodic the corresponding fermionic number operators here are 
\begin{align}
\hat{N}_{R,f}&=\sum_{n=0}^\infty{\te \left(n+\frac{1}{2}-\frac{\gamma}{2}\right)} \beta_{n+}^\dagger \beta_{n+}+\sum_{n=0}^\infty{\te \left(n+\frac{1}{2}+\frac{\gamma}{2}\right)}\beta_{n-}^\dagger \beta_{n-}\ ,\no \\ \la{234}
\hat{N}_{L,f}&=\sum_{n=0}^\infty{\te \left(n+\frac{1}{2}-\frac{\gamma}{2}\right)}
\tilde{\beta}_{n-}^\dagger \tilde{\beta}_{n-}+\sum_{n=0}^\infty{\te \left(n+\frac{1}{2}+\frac{\gamma}{2}\right)}
\tilde{\beta}_{n+}^\dagger\tilde{\beta}_{n+} \ , 
\end{align}
and the twisted-sector normal ordering constant turns out to be ($\gamma\in[0,1]$)
\begin{equation}\la{235}
A(\gamma)=2\, \zeta(-1, \gamma)-8\,\zeta(-1,\ha +\ha \gamma)+6\,\zeta(-1,1)=-1+\gamma\ .
\end{equation}
If we take $R\to\infty$ we end up with the 
type 0 spectrum in flat Minkowski space. 
 $R\to0$ at $\xi=0$ leads to the T-dual type 0 theory in flat space. In the limit of 
 $R\to0$ at $\xi=1$ (discussed above in the 
 type II case) 
 for 
 each value of the winding number we find one sector of integer frequency 
 fermionic modes ($A=0$) and one sector of half-integer frequency fermionic modes ($A=-1$),
 which originate alternatingly from untwisted and twisted sector. The condition that $\hat \mm$ 
 should vanish to get a finite-mass state from \rf{226} 
 can always be 
 satisfied and we again get the flat-space spectrum of the 
 T-dual type 0 theory. 
 
Finally, let us comment on the corresponding torus partition functions. 
 In \cite{Russo:1995tj,Takayanagi:2001jj} the partition function of type IIB theory on Melvin background was given as
\begin{align}\la{236}
	Z_{{\rm II B}}=&\frac{V_7R}{(2\pi)^{10}\alpha'^4}\int_{\F_0}\frac{\dd^2 \tau}{\tau_2^5}\sum_{w,w'\in\mathbb{Z}}\frac{\abs{\vartheta_{11}(\ha \xi\chi;\tau)}^8}{\abs{\eta(\tau)}^{18}\, \abs{\vartheta_{11}(\xi\chi;\tau)}^2}\ {\exp}\Big(-\frac{\pi R^2}{\alpha'\tau_2}\chi\bar{\chi}\Big)\ , \\
	&\chi\equiv w'-\tau w \ , \qquad \qquad \bar \chi\equiv w'-\bar \tau w \ . 
	\end{align}
	$Z_{{\rm IIB}}$ vanishes at $\xi=0$ when supersymmetry is restored. 
	The functions $\vartheta_{ab}(z;\tau)$ (see \rf{255}) are quasi-periodic with respect to shifts $z\to z+1$ and $z\to z+\tau$ but shifting $z$ by 
	$\ha $ or $\ha \tau$ relates the functions $\{\vartheta_{00},\vartheta_{01},\vartheta_{10},\vartheta_{11}\}$ to each other. 
	As a result, for $\xi=1$ the numerator in \rf{236} splits into four separate contributions while the denominator becomes proportional to $\vartheta_{11}(0;\tau)$. The denominator vanishes, indicating the appearance of new continuous zero modes. Finally, in the $R \to0$ limit the exponential factor becomes 1 and we recover the flat-space type 0A partition function given in \rf{215}. 
	
	In the case of type 0B string theory in Melvin background we should find the following analog of both \rf{215} and \rf{236}
	\begin{align}\la{237}
	Z_{{\rm 0 B}}=&\frac{V_7R}{(2\pi)^{10}\alpha'^4}\int_{\F_0}\frac{\dd^2 \tau}{\tau_2^5}\sum_{w,w'\in\mathbb{Z}}\frac{\II_{0B}}{\abs{\eta(\tau)}^{18}\ \abs{\vartheta_{11}(\xi\chi;\tau)}^2}\ {\exp}\Big(-\frac{\pi R^2}{\alpha'\tau_2}\chi\bar{\chi}\Big) \ , \\
	 \II_{\rm 0B}=&\te \abs{\vartheta_{00}(\ha \xi\chi;\tau)}^8+\abs{\vartheta_{01}(\ha \xi\chi;\tau)}^8+\abs{\vartheta_{10}(\ha \xi\chi;\tau)}^8+\abs{\vartheta_{11}(\ha \xi\chi;\tau)}^8. \la{238}
	 \end{align}

\section{Type 0B string in pp-wave background with RR 5-form flux}\label{8D}

A non-trivial example of type 0 string theory in curved space where the 
use of GS formulation is crucial is the $D=10$ pp-wave supported by RR 5-form flux. This is the same background as in type IIB theory, i.e. 
 the Penrose limit of $AdS_5\cross S^5$ with 5-form flux \ci{Blau:2001ne} ($i=1, ..., 8$) 
 \begin{equation}\la{301}
\dd s^2= 2 \dd u \dd v - f^2 x_i^2 \dd u ^2+\dd x_i^2\ , \qquad \qquad 
F_{u1234}=F_{u5678}= 4f \ .
\end{equation}
The corresponding type IIB string action and its spectrum were found in \cite{Metsaev:2001bj,Berenstein:2002jq,Metsaev:2002re}.
 After fixing the light-cone gauge on $u$, the type IIB GS action is as in flat space \rf{217} but now 8 bosons and 8 fermions acquire mass 
 $\mu$ proportional to the flux or the curvature scale $f$ \foot{Here 
 and below 
 $p^u=p_v $ is the component of the 
 string momentum defined with the standard factor of string tension ${1\ov 2\pi \a'}$.}
\begin{equation}\la{302}
u= u_0 + \a' p^u \tau \ , \qquad \qquad \mu=\alpha'p^u f \ .
\end{equation}
 The resulting light-cone Hamiltonian $H$ can be written as \cite{Metsaev:2002re}\footnote{Here we suppress indices on the complex mode operators (8 real modes are described by 4 complex ones
 and the sum is over both positive and negative $n$).
 In \cite{Metsaev:2002re} the fermionic modes appear multiplying certain combinations of gamma matrices. Here we redefined $\beta_n$ and $\beta_n^\dagger$ to absorb these.
 We have also fixed the ambiguity of choosing a 
 Clifford vacuum for the fermionic zero modes.} 
\begin{equation}\la{304}
 H=- p^v = \frac{1}{\alpha'p^u} \sum_{n=-\infty}
 ^\infty\sqrt{n^2+\mu^2}\left(\alpha_n^\dagger\alpha_n+\tilde\alpha_n^\dagger\tilde \alpha_n+\beta_n^\dagger\beta_n+\tilde\beta_n^\dagger\tilde\beta_n\right)\ , 
\end{equation}
where the zero-mode part has harmonic-oscillator type spectrum. 
The effective mass operator is then given by ($p_v= p^u$, \ $p^u={p^y+p^t \ov \sqrt{2}}$, \ $p^v={p^y-p^t\ov \sqrt{2}}$,\ $E=p^t$)
\begin{equation}\la{305}
M^2= E^2- (p^y)^2= -2p^u p^v=2p^u\, H \ , 
\end{equation}
reducing to the one in flat space in the limit of $\mu=0$. 
Note that since the eigenvalues of $H$ are expressed in terms of $\mu$ which depends on $p^u= {1\ov \sqrt 2} (E+ p^y)$ 
one gets here a non-trivial dispersion relation for the energy $E$ as a function of $p^y$ (see also below).

One way to find the type 0B string spectrum in this background 
is to compactify $y={u+v \ov \sqrt{2}}$ 
on a circle, 
 introduce a Melvin twist in some 2-plane of the transverse $x$-space, find the resulting type IIB 
 spectrum and then apply the same limit ($\xi\equiv q R =1$, $R\to 0$) as in section \ref{Melvin}.
 This is discussed in Appendix \ref{B}. 
 A short-cut is to apply the $(-1)^F$ orbifolding procedure, i.e. to combine the bosonic part of the pp-wave spectrum of type IIB theory (untwisted sector) with a similar spectrum found by imposing antiperiodic 
 boundary 
 conditions on the GS fermions (twisted sector). 
 This procedure was used in \cite{Bigazzi:2002gw,Takayanagi:2002pi}.

To cover both the case of the untwisted and of the twisted sector
we may write the corresponding light-cone Hamiltonian as 
\begin{align}\la{307}
 H&=\frac{1}{\alpha'p^u}\left(\hat N_b+\hat N_f+A\right)\ , \\\la{308}
\hat N_b&=\sum_{n=0}^{\infty}\omega_{n} 
(\alpha_{n}^\dagger \alpha_{n} + \tilde{\alpha}_{n}^\dagger \tilde{\alpha}_{n})
+ \sum_{n=1}^{\infty}\omega_{n}( \alpha_{-n}^{\dagger} \alpha_{-n} + \tilde{\alpha}_{-n}^\dagger\tilde{\alpha}_{-n}) \ , \qquad\omega_n =\sqrt{n^2 + \mu^2} ,\\\la{309}
\hat N_f&=
\sum_{n=a}^\infty\omega_{n} ( \beta_{n}^\dagger \beta_{n} + \tilde{\beta}_{n}^\dagger\tilde{\beta}_{n}) 
+ \sum_{n=1-a}^\infty\omega_{n}(\beta_{-n}^\dagger \beta_{-n}+ \tilde{\beta}_{-n}^\dagger\tilde{\beta}_{-n}) 
\ , \qquad a=0, \ \ha \ . \end{align} 
The untwisted sector corresponds to the choice of the lower summation limit $a=0$ in \rf{309} 
and then the normal ordering constant $A$ vanishes.
We are to project out all fermionic states, i.e. to consider only states created by an even number of 
worldsheet fermion operators. 

The twisted sector corresponds to $a=\ha$ and here the normal ordering constant is a non-trivial
function of $\mu$ \foot{We use the 
	label 8 to indicate that this model has 8 copies of massive excitations corresponding to 8-dimensional transverse space in \rf{301}.}
\begin{equation}\la{310}
A\equiv A_8(\mu) = 8\times \ha \Big(\sum_{n=0, \pm 1, ...}\sqrt{n^2 + \mu^2} - \sum_{r=\pm {1\ov 2}, ...} \sqrt{r^2 + \mu^2}\, \Big)\ . 
\end{equation}
It determines the mass of the ground state in the twisted sector. 
In the flat-space limit ($\mu=0$) we get\foot{The sum in \rf{310} for $\m=0$, i.e.
$ \ha \sum_{n=0, \pm 1, ...}|n| - \ha \sum_{r=\pm {1\ov 2}, ...} |r| $ 
 can be computed explicitly as:

$ \sum_{n=1}^{\infty} n - \sum_{n=1}^\infty ( n- \ha) \to \sum_{n=1}^\infty \big[ n \, e^{-\eps n} - (n-\ha) \, e^{-\eps (n-1/2) }\big] 
= - \ha e^{-\eps/2} ( 1 + e^{-\eps/2})^{-2} = -{1\ov 8} + \OO(\eps^2)$.} 
\be A_8(0) = 8 \, \big[ \zeta(-1,0) - \zeta(-1,\ha)\big] =-1 \ , \la{3100}
\ee 
determining the flat-space value of the type 0 tachyon mass 
 ($m^2_0 ={2\ov \a'} A(0)= - {2\ov \a'}$).

Separating the flat-space value and the bosonic $n=0$ term 
 we can represent \rf{310} as
\begin{align}\la{311}
A_8 &=-1+4\abs{\mu} + 8{\AA}(\mu) \ , \\\la{312}
\AA &\equiv\sum_{n=1}^{\infty}\Big[\sqrt{n^2 + \mu^2}-n - \sqrt{\left(n-\ha\right)^2 + \m^2 }+\left(n-\ha\right)\Big]\ . 
\end{align}
Expanding the square roots in \rf{312} 
 in powers of $\mu$ we can write $\AA$ as\foot{In \cite{Bigazzi:2002gw} \rf{310} was expressed in terms of the Epstein function. Here we will obtain a more convenient integral representation.} 
\begin{align}
\AA 
&=\sum_{n=1}^{\infty}\sum_{k=1}^\infty \tfrac{\mu^{2k} (-1)^k\Gamma(2k+1)}{2(1-2k)\left(\Gamma(k+1)\right)^2}\ \big[\tfrac{1}{(2n)^{2k-1}}-\tfrac{1}{\left(2n-1\right)^{2k-1}}\big]
=\sum_{k=1}^\infty \tfrac{\mu^{2k}\, (-1)^k \, \Gamma(2k-1)}{\Gamma(k)\Gamma(k+1)}\ \eta_{_D}(2k-1)\ , \la{313}
\end{align}
where we expressed the sum over $n$ in terms of the Dirichlet $\eta$-function
\begin{equation}\la{316}
\eta_{_D}(s)= \sum_{n=1}^\infty {(-1)^{n-1} \ov n^s} = ( 1 - 2^{1-s}) \zeta(s) =
\frac{1}{\Gamma(s)}\int_{0}^{\infty}\dd x~ \frac{x^{s-1}}{e^x+1}\ .
\end{equation}
The sum in \rf{313} converges for $\mu<\ha $. 
Analytic continuation to all $\mu$ can be achieved by using the integral representation of $\eta_{_D}$ thus getting 
\begin{align}
 A_8= -1 + 4 |\m| - 8 \abs{\mu}\int_{0}^{\infty}\dd x~ \frac{J_1(2 |\m| x)}{x\, (e^{x}+1)} \ ,
 \la{318}\end{align}
 where $J_1$ is the Bessel function of first kind.\foot{We used the following representation for the Bessel function:\ 
$J_\alpha(x)=\sum_{k=0}^{\infty}\frac{(-1)^k}{\Gamma(k+1)\Gamma(k+1+\alpha)}\left(\frac{x}{2}\right)^{2k+\alpha}
$.}
The resulting function $A_8$ in \rf{311} has the following asymptotics 
\begin{align}\la{319}
A_8\rvert_{\mu \to 0}=&-1+4\abs{\mu}-8\ln(2) \, \mu^2 + 6\, \zeta(3)\, \mu^4 - 15\, \zeta(5)\, \mu^6 +\order{\mu^8}\ , \\\la{320}
A_8\rvert_{\mu \to \infty}=&\, 0+\order{e^{- |\mu|}} \ , 
\end{align}
and its plot is given in Figure \ref{pic1}.
Note that the coefficient $\ln 2$ of $\mu^2$ in \rf{319} is the value of $\eta_{_D}(1)= \sum_{n=1}^\infty {(-1)^{n-1} \ov n} $. Also, 
 $A_8\rvert_{\mu \to \infty}\to 0$ has to do with the fact that 
the difference between the integer and half-integer mode contributions to \rf{301} disappears
 in the large $\mu$ limit.

Thus $A_8$ is asymptotically approaching zero from below, i.e. the ground state is ``tachyonic"
for all finite values of $\mu$, 
 becoming massless only at $\mu=\infty$.\foot{In the limit 
of $\mu \to \infty$ $A_8$ receives only exponential corrections
 (as was already mentioned in \ci{Bigazzi:2002gw}). The 
 structure of the exponential corrections in the integral in \rf{318} is similar to the one 
discussed in \ci{Schafer-Nameki:2005aui} and in \ci{Beccaria:2021ism}.} 
An important conclusion is that increasing the value of the RR flux makes the twisted-sector 
ground state less tachyonic. 

\begin{center}
	\includegraphics[width=10cm]{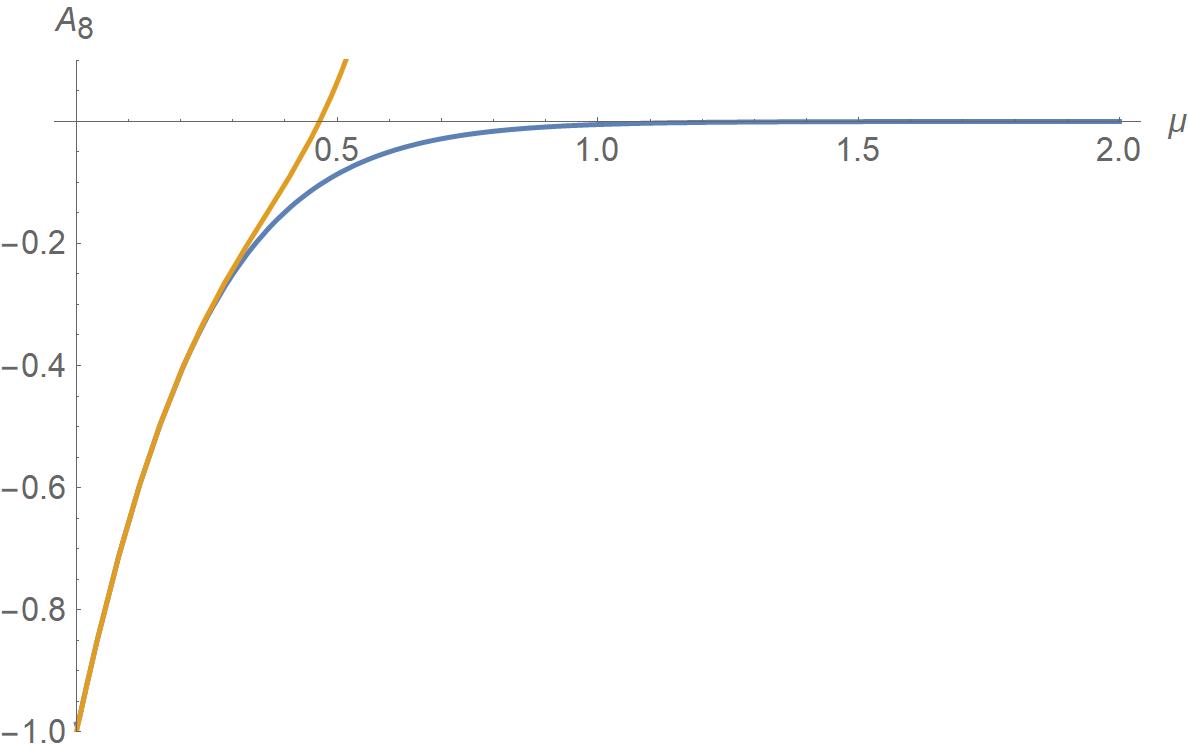}
	\captionof{figure}{\small {\small Dependence of the twisted-sector normal ordering constant $A_8$ on the 
	parameter $\mu=\a' p^u f $. The curve approaching zero at large $\m$ is obtained by numerical
	integration of \rf{318}, while the upper 
	 curve represents the small $\mu$ 
	 approximation \rf{319}}.}
 	\label{pic1}
\end{center}

To summarise, the untwisted sector of type 0 theory consists of the bosonic states of the 
type IIB spectrum. 
For example, the lowest energy states are given by even excitations of $\beta_0$ which are 
 128 bosonic states with masses $0$, $4\mu$, $8\mu$, $12\mu$ and $16\mu$. 
 The twisted sector has a unique tachyonic ground state with the effective mass $m^2= {2\ov \a'} A_8(\mu)$. 
 Both sectors contain also towers of massive states which are all bosonic.

Let us now comment on the meaning of the negative mass-squared of a field in the pp-wave background. 
Since the eigenvalues of $H$ in \rf{305} or \rf{307} are expressed in terms of $\mu$ which depends on $p^u= {1\ov \sqrt 2} (p^y+E)$ 
one gets a non-trivial ``dispersion relation'' for the energy $E=E(p^y)$ and one may wonder if the condition for stability may change compared to that in flat space. To see this let us start 
 with a simple massive scalar equation in the pp-wave metric \rf{301}
\begin{equation}\la{325}
 (\nabla^2 - m_0^2) T= (2\partial_u\partial_v+f^2x_i^2\partial_v^2+\partial_i^2-m_0^2) T=0 \ . 
\end{equation}
After the Fourier transformation in $u,v$ this becomes the equation for the wave function 
of the quantum harmonic oscillator with frequency $2f p_v$. Thus 
we get the following relation for $E=p^t$ as a function of $p^y$ \ ($p^v={p^y-p^t \ov \sqrt{2}}$, \ $p^u={p^y+p^t\ov \sqrt{2}}$)
\begin{align}\la{366}
-2p^u p^v & = m_0^2 + 2f p^uN_0 \ , \ \ \ \ \ \ \qquad N_0 = \sum_{i=1}^8(n_i+\ha )\ , \\
(p^t)^2 - (p^y)^2 &= m_0^2 + \sqrt 2 f (p^t+p^y) N_0 \ , \qquad \quad
E=p^t=\tfrac{fN_0}{\sqrt{2}}\pm\sqrt{(p^y+\tfrac{fN_0}{\sqrt{2}})^2+m_0^2}\ . \la{326} 
\end{align}
This implies that if $m_0^2$ 
is negative there is always a region of real momentum $p^y \sim -{fN_0\ov \sqrt{2}}$ 
for which the energy becomes imaginary, signalling an instability, i.e. the existence of solutions growing indefinitely 
with time (see also a discussion in \cite{Bigazzi:2003jk}). 

In the case of the string-theory tachyon in a non-trivial background, its equation receives $\a'$ corrections that
in the pp-wave case translate into its mass $m$ being itself a function of $\mu= \a' p^u f$. For 
 the type 0 spectrum discussed above this implies that 
 \be m^2_0 = -\te {2\ov \a'}\ \ \to \ \ m^2 = \te {2\ov \a'} A_8 (\a' p^u f ) = m_0^2 
 + 8 p^u f - 16 \ln (2)\ \a' (p^u f)^2 + \OO(\a'^2) \ , \la{344}
 \ee
 with $A_8$ given 
 in \rf{311},\rf{319}. 
From \rf{305} and \rf{307} we then learn that for the lowest excitation mode
we get 
\be \la{345}
E^2 - (p^y)^2 = \tfrac{2}{\alpha'} A_8\Big( \tfrac{\a' }{\sqrt{2}} ({p^y+E})f\Big ) \ . 
\ee 
Note that the $\a'$-independent term $8p^u f$ for \rf{344}
corresponds to the $ 2 f p^u N_0$ term in \rf{366}
in the oscillator ground state case of $n_i=0$, i.e. with $N_0=4$.

Eq. \rf{345} for $E(p^y)$ should be defined at any point of the real axis of $p^y$. 
Since $A_8 \leq0$ (cf. \rf{320} and Figure \ref{pic1}) we conclude that for small enough $(p^y)^2$ the value of $E$ is always imaginary, implying an instability. Note that 
the region of $p^y$ that leads to instability effectively shrinks with increasing $f$ as then $A_8$ grows towards zero.

The torus partition function corresponding to type 0B string theory 
in the above pp wave background 
was found in \cite{Takayanagi:2002pi} and 
can be written as\footnote{Ref. \cite{Takayanagi:2002pi} used different notation: the parameter $f$ was called $\mu$. 
Similar partition functions for type IIB theory in pp-wave background 
 at finite temperature were discussed in \cite{PandoZayas:2002hh, Greene:2002cd, Bigazzi:2003jk}.} 
\begin{align}\la{321}
Z_{\rm 0B} (f)&=\int\frac{\dd^2\tau}{2\tau_2}\, \int\dd p^u\dd p^v~e^{-2\pi\alpha'\tau_2\, p^up^v}\ \II(\tau; \a' p^u f) \ , \\
 \II(\tau;\mu)&= \frac{(\Z_{0,0})^4+ (\Z_{1/2,0})^4+(\Z_{0,1/2})^4+(\Z_{1/2,1/2})^4}{(\Z_{0,0})^4}\ , \la{333} \\
\no 
\Z_{a,b}(\tau;\mu)&=e^{4\pi\tau_2\Delta_b(\mu)}\prod_{n\in\mathbb{Z}}\left(1-e^{-2\pi\tau_2\abs{\omega_{n+b}}+2\pi i\tau_1(n+b)+2\pi i a}\right)\left(1-e^{-2\pi\tau_2\abs{\omega_{n-b}}+2\pi i\tau_1(n-b)-2\pi i a}\right) , \\\la{323}
\Delta_b(\mu)& = \sum_{k=1}^\infty c_k(\m) \, \cos(2\pi b\, k) \ , \qquad \qquad c_k= -\tfrac{1}{2\pi^2} \int_0^\infty\dd s~ e^{-k^2 s-\frac{\pi^2\mu^2}{s}} \ .
\end{align} 
Here $\omega_n$ was defined in \rf{308} and $c_k$ can be expressed in terms of the 
$K_1$ Bessel function. don't know this integral rep. -- check mathematica
 $\Z_{0,0}^4+\Z_{1/2,0}^4$ in \rf{333} 
correspond to the contribution of the untwisted sector and 
$\Z_{0,1/2}^4+\Z_{1/2,1/2}^4$ to that of the twisted sector. 

The structure of \rf{321} is very similar to that of the flat space partition function \rf{215}, with $\vartheta_{ab}\bar\vartheta_{ab}$ 
replaced by 
$\Z_{\frac{1-b}{2},\frac{1-a}{2}}$ (which indeed reduces to $\vartheta_{ab}\bar\vartheta_{ab}$ 
in the flat space limit $\mu\to0$). 
Note that in the type IIB case the pp-wave partition function has a similar structure but with the integrand 
being proportional to $\Z^4_{0,0}/\Z^4_{0,0}=1$, which mirrors the flat space expression \rf{219}.\foot{The type IIB pp-wave partition function does not automatically vanish, which is due to the zero modes lifting the flat-space degeneracy of the ground state so that instead of a full supermultiplet only one state remains massless \cite{Takayanagi:2002pi}. In the limit $\mu\to 0$ special attention has to be paid to these zero modes and one gets the vanishing flat space result \cite{Hammou:2002bf}. At the same time, 
one can also argue that the type IIB pp-wave partition function 
actually vanishes provided one uses an analytic continuation in the momentum integral in \rf{321}: integrating over the imaginary $p^v$ sets $p^u$ and thus $\m$ to zero and one recovers the flat-space result $Z=0$. } 

Since the partition function \rf{321} encodes the type 0 spectrum, one is able to recover from it the value of the effective mass of the ground state. Taking the limit $\tau_2 \to \infty$
the ground state contribution to $\Z_{a,b}$ comes from the exponential $e^{4\pi\tau_2\Delta_b {(\mu)}}$ and 
thus the twisted sector contribution gives\foot{In the untwisted sector similar 
 exponentials cancel corresponding to a massless lowest-energy state.} 
\begin{equation}\la{324}
\frac{\Z_{0,1/2}^4+\Z_{1/2,1/2}^4}{\Z_{0,0}^4}\ \to \ e^{16\pi\tau_2[\Delta_{1/2}(\m) -\Delta_0(\m) ] }=e^{-2\pi\tau_2A_8}\ , 
\end{equation}
 where $A_8$ is the same as in \rf{311}. 

\section{Type 0B string in pp-wave background with mixed \\ NSNS and RR 3-form fluxes}\label{4D}

Let us now consider type 0B theory 
in the pp-wave background which is a Penrose limit of $AdS_3\times S^3\times \rT^4 $
supported by a mixture of NSNS and RR 3-form fluxes. This background is a solution of both type IIB and type 0B 
theories and thus the type 0B spectrum can be found again using the GS description and applying the $(-1)^F$ orbifolding procedure.
The explicit form of the metric and fluxes is ($r=1,2,3,4; \ s=5,6,7,8$) 
\begin{equation}\la{401}
\begin{split}
\dd s^2&= 2 \dd u \dd v - f^2 x_r^2 \dd u ^2+\dd x_r^2 + \dd \xx_s^2 \ , \\
H_{u12}&=H_{u34}=-2qf,\qquad\qquad F_{u12}=F_{u34}=-2\sqrt{1-q^2}f\ , 
\end{split}
\end{equation}
where $\xx_s$ are coordinates of the 4-torus $\rT^4$. The parameter $q$ (that we shall assume to take values in the interval $[0, 1]$) interpolates between the pure RR case ($q=0$) 
and the pure NSNS case ($q=1$). The type IIB string in this background and its spectrum was discussed in \cite{Berenstein:2002jq,Russo:2002rq,Grignani:2003cs} (see also \ci{Cagnazzo:2012se,Hoare:2013pma}).

The light-cone Hamiltonian is given again by \rf{307}, i.e. 
\begin{equation}\la{402}
H=\frac{1}{\alpha'p^u}\left(\hat N_b+\hat N_f+\hat N_{T}+A\right)\ ,
\end{equation}
where $\hat N_b$ and $\hat N_f$ 
have the same form as in \rf{308}\rf{309}
 but now with half of the 
 massive bosonic and fermionic oscillator modes and the frequencies given by
\begin{equation}\la{403}
\omega_n=\sqrt{(n+q\, \mu)^2 + (1-q^2)\, \mu^2} \ , \qquad \qquad \mu = \a' p^u f \ . 
\end{equation}
$\hat N_\rT$ stands for the contribution of the 4 massless torus bosons and 4 massless decoupled fermions
(there is also the standard $\rT^4$ momentum and winding mode contribution that we suppress). 
The NSNS flux acts to shift the mode number $n$ (introducing certain periodicity, see below) while RR flux produces the 
effective mass $\mu' =\sqrt{1-q^2}\mu$. 

The normal ordering constant $A$ vanishes in the 
type IIB case and thus in the untwisted sector of the 
type 0B spectrum, 
while in the twisted sector where GS fermions are taken to be antiperiodic 
we get (cf. \rf{310},\rf{3100})
\begin{align}\la{404}
&A\equiv A_4(\mu, q)=A_\rT+\bA_4(\mu, q)\ , 
\quad \quad \quad A_\rT =\bA_4(0,q) = 4\, \big[ \zeta(-1,0) - \zeta( -1,\ha) \big]= -\tfrac{1}{2}\ ,
\end{align}
\begin{align}\la{405} 
&\bA_4(\mu,q)=4 \times \ha \Big[ \sum_{n=0, \pm 1, ...}\sqrt{(n+q\, \mu)^2 + (1-q^2)\mu^2}
- \sum_{r=\pm {1\ov 2}, ...} \sqrt{( r + q\, \mu)^2 + (1-q^2)\mu^2} \Big]. \end{align}
 $A_\rT$ in \rf{404} is the contribution of massless oscillators in $\hat N_\rT$. In the limit of $\mu=0$ we recover the flat space value $A_4 = A_\rT+\bA_4(0, q) =-1$ as in \rf{3100}. 

Separating the flat-space part and the $n=0$ bosonic contribution in 
the sum in \rf{405} as in \rf{311},\rf{312}, the function $A_4$ can be represented as 
\begin{align}\la{4044}
& A_4 (\mu, q) = -\ha + \bA_4(\mu,q) = - 1 + 2 |\mu| 
+ 2\sum_{n=1}^{\infty} X_n \ , \\
&X_n= \sqrt{(n+q\, \mu)^2 + (1-q^2)\mu^2} + \sqrt{(n-q\, \mu)^2 + (1-q^2)\mu^2} \no \\ 
&\qquad 
- \sqrt{( n-\tfrac{1}{2}+ q\, \mu)^2 + (1-q^2)\mu^2} - \sqrt{( n-\tfrac{1}{2}- q\, \mu)^2 + (1-q^2)\mu^2}\ -1 \ . \la{57}
\end{align}
For $n \gg 1$ one has $X_n \to -
 \frac{\m^2 (1-q^2) }{2 n^2}+\OO( {1\ov n^3})$, 
 i.e. the sum in \rf{4044} is convergent. 

In the pure RR flux case of $q=0$ we get (cf. \rf{310} and \rf{405})
\begin{equation}\la{406}
A_4(\mu,0)=- \ha + \ha {A_{8}(\mu)} \ , 
\end{equation}
i.e. $A_4(\mu,0)$ is always negative, growing to $-\ha $ at $\mu\to \infty$. 

In the pure NSNS flux case of $q=1$ \rf{4044} simplifies to 
\begin{equation}\la{407}
A_{4}(\mu, 1) =- 1 + 2 |\m| + 2\sum_{n=1}^{\infty} \Big( | n + \mu| + | n - \mu| - |n-\tfrac{1}{2}+ \mu|
- |n- \tfrac{1}{2}- \mu|\ -1 \Big) \ .
\end{equation}
For $0 \leq \mu \leq \ha$ we find that $\bA_{4}(\mu, 1) = 2 \mu -1 $ and for general $\mu$ this function 
 oscillates between -1 at $|\mu|= k$ and 0 at $|\mu|= k + \ha $, $k=0, 1, 2, ...$ with period 1
(see Figure \ref{pic2}).\foot{This is similar to the periodicity in the normal ordering constant \rf{231} 
 in the Melvin background discussed in section \ref{Melvin},
 where a renaming of mode numbers allowed to absorb integer shifts of the parameter $\gamma$ in \rf{225}.}
 Thus the normal ordering constant vanishes for special values of $\mu$.
To analyse stability we need to fix the background parameter $f$; 
since $p^u$ may vary, we will still find an 
instability at other values of $\mu=\a' p^u f$. 
 
\begin{center}
	\includegraphics[width=7cm]{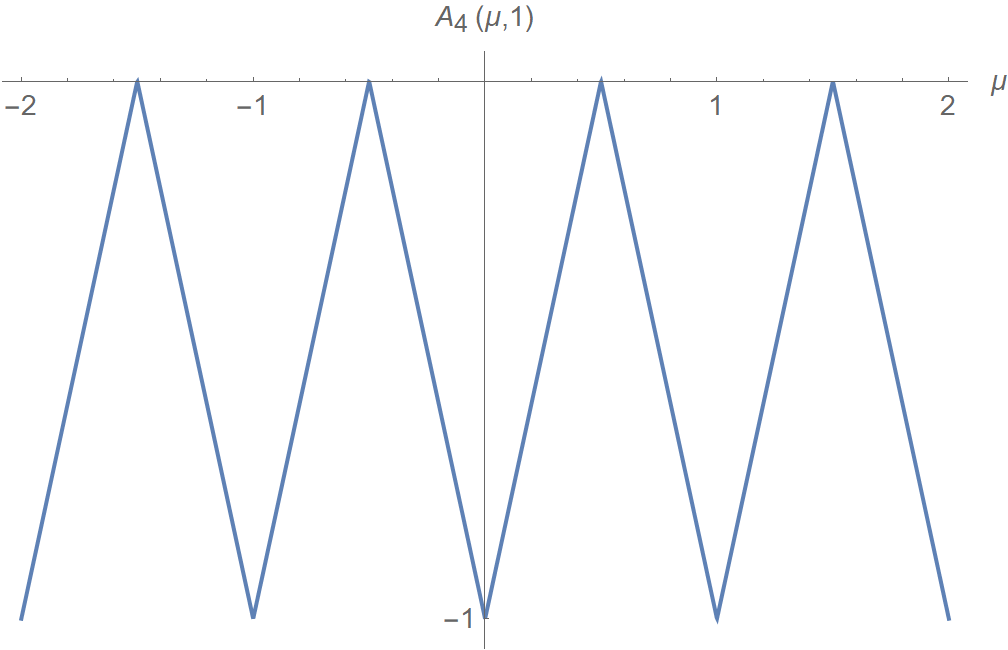} 
	\captionof{figure}{\small Plot of $A_4$ as a function of $\mu$ in the $q=1$ case (pure NSNS flux).}
	\label{pic2}
\end{center}

For $q \in (0,1) $ the asymptotics of $A_4=-\ha + \bA_4$ are given by (cf. \rf{319},\rf{320})
\begin{align}\la{410}
&A_4\rvert_{\mu \to 0}=-1+2\abs{\mu}-4\ln(2)(1-q^2)\, \mu^2 +3\zeta(3)(1-q^2)(1-5q^2)\, \mu^4 +\order{\mu^6}\ ,\\\la{411}
&A_4\rvert_{\mu \to \infty}=-\tfrac{1}{2}+\order{e^{-|\mu|}}\ . 
\end{align}
 For general $q$ one can show that $\bA_4$ vanishes at special values of $\mu$ where $ 4 q\m$=odd integer, i.e. 
 \be \ A_4=-\ha +\bA_4= -\ha \ \quad \ \ \ {\rm if} \ \ \ \ q\, \mu = \tfrac{1}{ 4} (2k+1)\ , \ \quad \ k \in \mathbb{Z} \ . \la{400}\ee
 This follows from rearranging terms in the convergent sum in $\bA_4$ in \rf{4044}. For example, 
 for $q\mu = {1\ov 4}$ the second and the third terms in \rf{57} cancel and then including the zero-mode term 
 $2 |\mu|= \sqrt{ (q \mu )^2 + (1- q^2) \mu^2} $ in the sum of the first term, in \rf{57} and renaming $n$ in the sum of the fourth term, one concludes that $\bA_4\big|_{q\mu={1\ov 4}}= 0$. 

 One can find an integral representation for $A_4$ similar to the one for $A_8$ in \rf{318} by considering the intervals 
 between the points $q\mu = \four ( 2 k+1)$ separately. 
 For example, 
 for $q \mu \in [- \frac{1}{4},\frac{1}{4}]$ we get 
\begin{equation}\la{409}
A_4
=-\tfrac{1}{2}+2 \sqrt{(q \mu)^2 + \m'^2}
-4 |{\mu' }|\int_{0}^{\infty}\dd x~ \frac{ J_1(2|\mu' |\, x)} {x\, (e^{x}+1)}\ \cosh (2 q \mu\, x)\ , \qquad \m' \equiv \sqrt{1-q^2} \, \mu \ . 
\end{equation}
To get the analog of \rf{409} 
for $q\mu \in [\frac{1}{4},\frac{3}{4 }]$ we need to introduce an overall minus sign for the $\bA_4$ contribution 
and shift $\mu$ by $\frac{1}{2q}$ while leaving $\mu' $ unaltered. 
$A_4$ is continuous at the 
 glueing points $q\mu = \four ( 2 k+1)$ where it takes the value $A_4=-\ha$. 
The plots of $A_4$ as a function of $\mu$ for some values of $q$ are shown in Figure \ref{pic3}. 

\begin{center}
	\includegraphics[width=10cm]{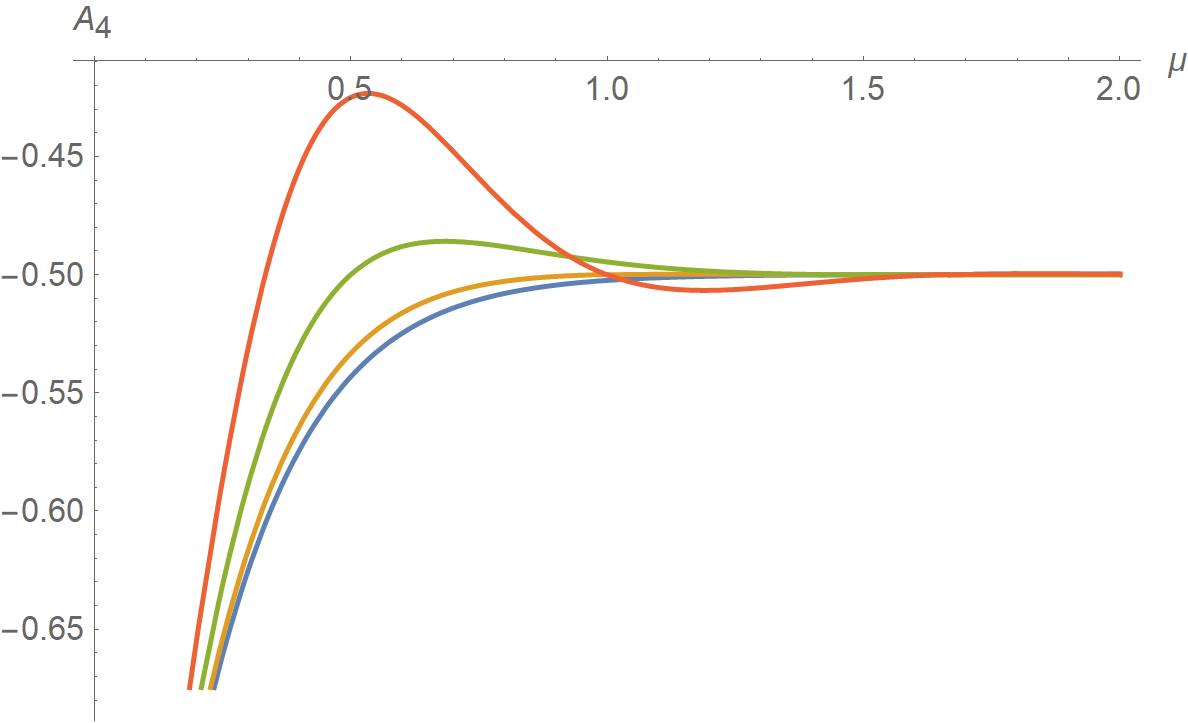}
	\captionof{figure}{Dependence of the normal ordering constant $A_4$ on the parameter $\mu$ for different values of $q$: 
	 $\frac{3}{4}$ (upper red curve), $\frac{1}{2}$, $\frac{1}{4}$, and $0$ (lower blue curve). Note that
	 $A_4=-\ha$ for $q \mu =\ha k + { 1 \ov 4}, \ k\in\mathbb{Z} $.} 
 	\label{pic3}
\end{center} 



Let us now consider the corresponding type 0 partition function, i.e. the analog of \rf{321}. 
We will need to include the torus $\rT^4$ contribution, i.e. 
factors of the massless oscillator partition function in general defined as 
\begin{align}
\Z_{a,b}^\rT(\tau)=&e^{4\pi\tau_2\left[\frac{1}{24}-\frac{1}{8}\left(2b-1\right)^2\right]}\, \prod_{n\in\mathbb{Z}}\left(1-e^{-2\pi\tau_2\abs{n+b}+2\pi i\tau_1(n+b)+2\pi i a}\right)\no \\
& \quad \qquad \qquad \qquad \times \left(1-e^{-2\pi\tau_2\abs{n-b}+2\pi i\tau_1(n-b)-2\pi i a}\right)
= \abs{\eta(\tau)}^{-2 } \, \Big|\theta{\tiny \begin{bmatrix} {\ha} -b\\{\ha} -a\end{bmatrix}}(0,\tau)\Big|^2 \ , \la{413}
\end{align}
where $b\in[0,1]$ and the 
$\theta$-function is 
defined in \rf{255}. The bosonic $\rT^4$ momentum and winding 
 modes contribute
\begin{equation}\la{414}
\Z_0(\tau) =\Tr\, e^{i\pi (\tau p_R^2- \bar{\tau}p_L^2)} =\sum_{p_L, p_R }e^{i\pi\tau_1(p_R^2-p_L^2)}e^{-\pi\tau_2(p_R^2+p_L^2)} \ , 
\end{equation}
where for the simplest rectangular torus with $R=\sqrt{\a'}$ one has 
$(p_{L, R})_s = \mm_s \pm w_s$. 
This factor replaces the contribution $\tau_2^{-4/2}$ of 4 continuous bosonic zero modes. 
Let us also introduce 
\begin{equation}\la{412}
\tZ_{a,b}(\tau;\mu,q)=\Z_{a-q\mu\tau,\, b+q\mu}(\tau;\sqrt{1-q^2}\,\mu) \ , 
\end{equation}
where $Z_{a,b}(\tau;\mu)$ was defined in \rf{333}
Then the total partition function may be represented as (cf. \rf{321}) 
\begin{align}\la{415}
&Z_{\rm 0B}(f,q)=\int \frac{\dd^2\tau}{2\tau_2}\int\dd p^u\dd p^v~e^{-2\pi\alpha'\tau_2p^up^v}\, \Z_0(\tau) \ \II(\tau; \a' p^u f, \, q ) \ , \\
& \II= \frac{(\tZ_{0,0})^2(\Z^{\rT}_{0,0})^2+(\tZ_{1/2,0})^2(\Z^{\rT}_{1/2,0})^2+(\tZ_{0,1/2})^2(\Z^{\rT}_{0,1/2})^2+(\tZ_{1/2,1/2})^2(\Z^{\rT}_{1/2,1/2})^2}{(\tZ_{0,0})^2\ \abs{\eta(\tau)}^8} \ , \la{4165}
\end{align}
where in \rf{4165} we suppressed the arguments of the $\Z$-functions. 
Modular invariance can be checked along the same lines as for \rf{321} (see \cite{Takayanagi:2002pi}).
 A few useful relations are 
\begin{align}\la{416}
&\tZ_{a,b}(\tau+1;\mu,q)=\tZ_{a+b,b}(\tau;\mu,q),\qquad \tZ_{a+1,b}(\tau;\mu,q)=\tZ_{a,b+1}(\tau;\mu,q)=\tZ_{a,b}(\tau;\mu,q),\\\la{417}
&\Z_{a,b}^\rT(\tau+1)=\Z_{a+b,b}^T(\tau),\qquad \Z_{a+1,b}^T(\tau)=\Z_{a,b+1}^\rT(\tau)=\Z_{a,b}^\rT(\tau) \, \qquad 
\Z_{a,b}^\rT\left(-\tfrac{1}{\tau}\right)=\Z_{-b,a}^\rT(\tau). 
\end{align} 
Also, we may use that \cite{Takayanagi:2002pi} 
\begin{equation}\la{419}
\Z_{a,b}\left(-\tfrac{1}{\tau}; \abs{\tau}\mu\right)=\Z_{-b,a}(\tau;\mu)\ .
\end{equation}
In the present case the periodicities transform with $\mu$ as a modular parameter rather than a positive constant,
 so we have to be careful about where the absolute value should be taken. 
 Following a similar discussion in \cite{Sugawara:2002rs},
 one can show that 
\begin{equation}\la{420}
\tZ_{a,b}\left(-\tfrac{1}{\tau};\tau\mu,q\right)=\Z_{a+q\mu, b+q\mu\tau}\big(-\tfrac{1}{\tau};\tau\sqrt{1-q^2}\mu\big)=\Z_{-b-q\mu\tau, a+q\mu}(\tau;\sqrt{1-q^2}\mu)=\tZ_{-b,a}(\tau;\mu,q).
\end{equation}
As a result, one can check that 
the integrand in \rf{415} is invariant under $\tau\to-{1\ov \tau}$ provided we also redefine 
the integration variables $p^u\to\tau p^u$,\ $p^v\to\bar{\tau}p^v$.

Taking the large $\tau_2$ limit one finds as in \rf{324} that it is controlled by the mass 
of the ground state in the twisted sector, i.e. by the normal ordering constant $A_4$ in \rf{404}. Namely, we get 
\begin{equation}\la{421}
\exp( 2\pi\tau_2\Big[2\Delta_{{1\ov 2} +q\mu} (\sqrt{1-q^2}\mu) -2\Delta_{q\mu } (\sqrt{1-q^2}\mu) +\tfrac{1}{2} \Big])\ , 
\end{equation}
where $\Delta_b (\mu)$ was defined in \rf{323}. As follows from 
 \rf{323}, this expression exactly matches the normal ordering constant \rf{409}.
Since $A_4$ is negative we find as in \rf{345} that there is always a region of instability. 

\section{Remarks on the \adss case} 
As we discussed above, in flat space and in the light cone gauge the NSR description of type 0 string theory and its GS description based on the orbifolding procedure are equivalent \cite{Dixon:1986iz,Klebanov:1999pw}. This generalises to solvable examples of curved NS-NS models like pp-wave and Melvin-type background. The case of RR backgrounds is non-trivial and to argue that the same orbifolding procedure applies we studied the simplest example of pp-wave supported by RR flux. 
 The corresponding spectrum of type 0 string theory in pp-wave background 
 supported by $F_5$ flux 
 was already found  using the orbifolding procedure \cite{Bigazzi:2002gw,Takayanagi:2002pi}, 
 while we demonstrated (in Appendix \ref{B}) that like in flat space 
 the reasoning for this can be understood by starting with the type IIB case, adding a Melvin twist and taking the special limit that leads to the type 0 string.
 
Let us now comment on the example of our main interest -- \adss background supported by self-dual
RR 5-form flux which is a 
 solution of both type IIB and type 0B theories \ci{Klebanov:1998yya}. 
The spectrum of the 
type 0B string in this background should again be 
given by a combination of the 
untwisted and twisted sectors
described by the \adss GS action with periodic and antiperiodic fermions
respectively. 
Indeed, the above Melvin-twist procedure leading to an orbifolding construction 
 of the spectrum in an RR-flux supported pp-wave background 
 should apply also to the \adss case. However, in 
 this case the type IIB spectrum (with Melvin twist added) is not known 
 in an explicit form and thus one may not be able 
 to directly implement the limiting procedure leading to the type 0 spectrum. 
 Thus this orbifolding prescription may still be viewed as a conjecture. 
 We believe it is a natural one, 
 given that the corresponding dual gauge theory may be interpreted as an orbifold of $\N=4$ SYM theory \ci{Klebanov:1999ch,Nekrasov:1999mn}.

The classical type IIB GS string in \adss 
 background \ci{Metsaev:1998it} is known to be integrable \ci{Bena:2003wd} and by now its spectrum 
 is effectively known, at least implicitly \ci{Beisert:2010jr,Gromov:2014caa}. 
 The integrability property should be expected to determine 
 the spectrum in the twisted sector as well since 
 it should also 
 apply to the case of antiperiodic GS fermions.

\def \ks {{\rm s}} 

To get an idea of the effect the choice of fermion periodicity has, 
let us consider some semiclassical string states with large quantum numbers in the limit of large effective string tension ${\sql \ov 2\pi}={ R^2_{\rm AdS}\ov 2\pi \a'}$.
In the semiclassical limit \ci{Gubser:2002tv,Frolov:2002av,Tseytlin:2004xa,Tseytlin:2010jv} one fixes the parameters of the classical string solution or the ratios of spins to string tension while taking $ { \sql}\gg 1$, e.g., 
\be {S\ov \sql}\, , \ { J\ov \sql} ={\rm fixed} \ , \qquad \ \ \sql \gg 1 \ ,\la{511} \ee
where $S$ is a spin in $AdS_5$ and $J$ is an angular momentum in $S^5$. 
One important example is the long folded spinning string in $AdS_5$ with the energy \ci{Gubser:2002tv}
\begin{align}
&\qquad \qquad \qquad E= S + f(\l) \ln { S} + ... \ , \qquad \qquad \sql \gg 1, \quad { S\ov \sql } \gg 1 \ , \la{61}\\
& f(\l) = \tfrac{1}{ \pi} \big[ a_0 {\sql} + a_1 + \tfrac{ 1}{ \sql } a_2 + ...\big] \ ,\qquad \qquad a_0 = 1 
\ , \ \ \ 
a_1=- { 3 \ln 2 } \ , \ \ \ a_2 =- {\rm K} \ , ... \ . \la{62}
 \end{align}
The function 
$f(\l)$ is the cusp anomalous dimension which is known exactly from the integrability \ci{Beisert:2006ez,Basso:2007wd}.
In the semiclassical string expansion, 
$a_0$ is determined by the classical bosonic string solution \ci{Gubser:2002tv}, 
$a_1$ -- by the 1-loop correction due to bosonic plus fermionic 
 string fluctuations \ci{Frolov:2002av}, $a_2$ (proportional to Catalan's constant $K$) -- by the 2-loop GS string correction 
\ci{Roiban:2007jf}, etc. .
Explicitly, $a_1$ is given by the sum of contributions of the 8 bosonic and 8 fermionic fluctuation 
 modes (${\rm s}= {1\ov \pi}\ln {S\ov \sql} \gg 1$) \ci{Frolov:2002av}
 \begin{align} a_1= & \lim_{\ks \to \infty} {1\ov \ks^2} 
	\sum_{n=1}^\infty \big( \sqrt{n^2 + 4 \ks^2 } + 2 \sqrt{n^2 + 2 \ks^2}
+ 5 \sqrt{n^2} - 8 \sqrt{n^2 + \ks^2} \ \big)
\no \\ =& \int^\infty_0 dp\ \big( \sqrt{p^2 + 4 } + 2 \sqrt{p^2 + 2 }+ 5p - 8 \sqrt{p^2 + 1} \ \big) 
= - {3 \ln 2 }
 \ . \la{63} \end{align}
A similar semiclassical state 
should exist in the twisted sector of type 0B theory where the GS fermions are antiperiodic. 
Because of the limit ${S\ov \sql} \gg 1$ the folded string is infinitely long, i.e. the world sheet circle becomes effectively decompactified and the 
sign of periodicity of the 
fermions becomes irrelevant. Thus we should find the same value of the function $f(\l)$.
 Indeed, in the antiperiodic case the sum in \rf{63} is replaced by (cf. \rf{310})
\be 
\sum_{n=1}^\infty \Big[ \sqrt{n^2 + 4 \ks^2 } + 2 \sqrt{n^2 + 2 \ks^2}
+ 5 \sqrt{n^2} - 8 \sqrt{(n-\ha)^2 + \ks^2} \ \Big] \ . \ee 
The trivial $\ks$-independent divergent part in this sum drops out from $a_1$ for $\ks\to \infty$\foot{An alternative 
is to isolate the divergent $\ks=0$ part explicitly as in \rf{311},\rf{312} 
which will ``symmetrise" the fermionic contribution (cf. \ci{Frolov:2004bh}):
$8 \sqrt{(n-\ha)^2 + \ks^2} \to 4 \sqrt{(n-\ha)^2 + \ks^2} + 4 \sqrt{(n+\ha)^2 + \ks^2} $.}
and then setting $n \to p\, \ks$ we conclude that the 
$-\ha$ shift in the summation index in the fermionic contribution does not play a role in the large spin limit. 
Thus we arrive at the conclusion that the energy of the corresponding fast spinning string state in the twisted
sector should be the same as in the untwisted sector.

To get a non-trivial effect from the change of periodicity of the GS fermions one is to consider solutions 
where the size of the closed string remains fixed in units of $\sql$. For example, in the 
 limit where the $S^5$ angular momentum parameter $\J = {J \ov \sql}$ is taken to be larger than other 
 semiclassical string parameters 
 one should recover the pp-wave spectrum with the same twisted sector ground state energy as in \rf{310}. 
Another simple example is provided by the rigid circular string rotating in two orthogonal planes in the $S^3$ part of $S^5$ 
 with the two equal angular momenta $J_1=J_2 = \ha J $ \ci{Frolov:2003qc,Arutyunov:2003za}. 
 Its classical $AdS_5$ energy is given by 
\be
 E_0 = \sql\, \k = \sqrt{ J^2 + \l \rk^2}
 = J + \te { \l \rk^2 \ov 2 J} + ... \ , \la{cla} \ee
where $\rk$ is the winding number of the string and in the last equality we expanded in large $\J= {J\ov \sql}$.\foot{The large $\J$ limit is 
 similar to the expansion near the pp-wave background (cf. \rf{302},\rf{307}) with 
 $p^u \to J$ and $\mu = \a' p^u f 
 \to \J= {J\ov \sql }= \a' J R^{-2}_{\rm AdS}$.}

Starting with the form of this solution in \ci{Arutyunov:2003za}, one finds for the 1-loop correction to the energy 
$E=E_0 + E_1 + ... $ \ci{Beisert:2005mq,Minahan:2005qj}\foot{The fluctuation frequencies were found in \ci{Frolov:2003tu} starting with the form of the solution in \ci{Frolov:2003qc} related to the one in \ci{Arutyunov:2003za} by an $SO(4)$ transformation. 
 In \ci{Frolov:2004bh} it was assumed that after a rotation needed to diagonalise the fermionic fluctuation 
 Lagrangian the GS fermions become effectively antiperiodic. In fact, this rotation is not needed 
 if one starts with the form of the solution in \ci{Arutyunov:2003za}, i.e. the type IIB fermions remain periodic. 
 This issue was further clarified in \ci{Mikhaylov:2010ib}.}
\begin{align} 
 &E_1 = E_{1,0} + E'_1 \ , \ \ \ \ \ \ \ \ \ \
 E'_{1}= \sum^\infty_{n=1} S_n \ , \la{su} \\
 & E_{1,0}=\te \big[ 2 + \sqrt{1 -{ 2 \rk^2\ov \k^2 } } + 5 \sqrt{1 -{ \rk^2\ov \k^2} } \big] - 8 \sqrt{1 -{ \rk^2\ov \k^2} }
\ ,\la{dou} \qquad\qquad \k^2 = \J^2 + \rk^2 \ , \\ &
 S_n
 = \te 2 \sqrt{ 1 +
 { ( n + \sqrt{n^2 - 4 \rk^2 } )^2 \ov 4 \k^2 } }
+{2}\sqrt{1 + {n^2-2\rk^2\ov \k^2 } }
 +{4}\sqrt{ 1 + {n^2\ov \k^2 } } 
-{8}\ \sqrt{1 + { n^2-\rk^2 \ov \k^2 } }
 \ . \la{hi}
 \end{align} 
Here $ E_{1,0}$ is the contribution of the zero modes and the 
negative contributions come from 8 periodic fermionic fluctuations. 
For the corresponding state in the twisted sector of type 0B theory one needs to take the fermions to be antiperiodic, i.e. 
 drop the fermionic zero mode term in $E_{1,0}$ and sum the negative term in \rf{hi} with $n \to r= n -\ha $. 
Then the difference between the energies of the untwisted and twisted sector states is found to be 
\begin{align}
 \la{519}
\Delta E_1 = E_1- \tilde E_1= & - \tfrac{1}{ \k}\, 8 \times \ha \Big[ \sum^\infty_{n=0, \pm 1, ...} 
 \sqrt{ n^2 + \J^2} - \sum^\infty_{r=\pm {1\ov 2}, ...} \sqrt{ r^2 + \J^2} \Big]\no \\
 = & - {1 \ov \sqrt{\J^2 + \rk^2}} A_8 (\J) \ .
 \end{align}
Here $A_8(\J) $ is exactly the same function \rf{310},\rf{318}
 as in the ground state energy of the twisted sector 
of the type 0B theory in pp-wave background. 
A similar discussion can be repeated for another simple 
case of the circular $(S,J)$ solution \ci{Park:2005ji}.

One may also use a semiclassical approach to study quantum corrections to energies of short strings \ci{Beccaria:2012xm}
but addressing 
the question about the energy of the type 0B tachyon as a function of the string tension $\sql$ 
 requires an exact solution for the twisted sector of the 
\adss spectrum. Hopefully, this can be achieved using the integrability of the \adss GS string theory
that should hold regardless the choice of periodicity for the GS fermions.

 \iffa
 in our paper with Beisert and Hernandez-Lopez https://arxiv.org/pdf/hep-th/0603204.pdf
(S,J) energy was used to fix phase in scattering matrix. 
it should not depend on periodicity of fermions as is determined essentially on a plane rather than cylinder but could you check this carefully. Then statement is that regardless periodicity of fermions BMN S-matrix and phase are same as in type II B theory. 
Difference between twisted and untwisted sectors appears only once we consider theormodynamic Bethe ansatz that is needed to compute the spectrum... we should have added this remark in sect 5. i will create v3 for the future. 
 \fi


\section*{Acknowledgements}
AAT is grateful to M. Beccaria, F. Levkovich-Maslyuk, J. Russo and S. van Tongeren 
 for related discussions. We also thank R. Metsaev for comments on the draft. 
 TS acknowledges funding by the President's PhD Scholarships of Imperial College London. 
 The work of AAT was supported by the STFC grant ST/T000791/1. 

\

\appendix
\section{
Type 0 string in flat space}\label{A}

Here we shall recall the standard description of type 0 string in flat space first in the NSR and then in the light-cone GS approach. 

We shall start with NSR closed string theory on a cylinder $(\tau,\sigma)$ and use the notation $\sigma^\pm=\tau\pm\sigma$, \ $\partial_{\pm}=\frac{1}{2}(\partial_\tau\pm\partial_\sigma)$. The light-cone action is ($i=1, ..., 8$) 
\begin{equation}\la{201}
\S=\frac{1}{\pi\alpha'}\int\dd^2\sigma~\left(\partial_+x^i\partial_-x^i+i\bar{\psi}_+^i\partial_-\psi^i _{+}+i\bar{\psi}_-^i\partial_+\psi^i_{-}\right)~.
\end{equation}
Depending on periodicity of left- and right-moving fermions we get 4 sectors. 
In each left- or right-moving sector the mass operator is given by 
\begin{equation}\la{202}
\alpha'M^2 = \sum_{n=1}^{\infty}n\alpha^i_{-n}\alpha^i_n+\sum_{r=b}^{\infty}r\beta^i_{-r}\beta^i_r+A~,
\end{equation}
where $\alpha^i_{-n}(\alpha^i_{n})$, $\beta^i_{-n}(\beta^i_{n})$ are creation (annihilation) operators of bosonic and fermionic modes respectively. 
The values of the 
parameter $b$ in the fermionic sum and the normal ordering constant $A$ depend on the sector: 
in the NS (antiperiodic) sector $b=\ha $ and $A=-\ha $ while in the R (periodic) sector $b=1$ and $A=0$. 
The ground state $\ket{0}_{NS}$ is a scalar tachyon while $\ket{0}_R$ is an $SO(8)$ spinor. 
Combining the left- and right-moving 
sectors and imposing the level matching condition $M^2_L=M^2_R$
one then generates the closed string spectrum.
In particular, one gets\foot{
	One introduces a $\mathbb{Z}_2$ grading by defining ``G-parity" as $(-1)^{{\rm F}+1}$ for the NS-sector and $(-1)^{\rm F}\Gamma_9$ for the R sector (here $\rm F$ counts the number of worldsheet fermionic excitations).} 
\begin{center}
	\begin{tabular}{c | c | c | c}
		Sector & Lowest mass state & Rep of SO(8) & Statistics \\ 
		\hline
		(NS$_-$, NS$_-$) & $\ket{0}_{NS}\otimes \ket{0}_{NS}$ & $\mathbb{1}\otimes\mathbb{1}$ & Bosonic \\
		(NS$_+$, NS$_+$) & $\xi_{ij}\beta^i_{-1/2}\ket{0}_{NS}\otimes \beta^j_{-1/2}\ket{0}_{NS}$ & $8_v\otimes 8_v$ & Bosonic \\
		(R$_+$, R$_+$) & $\ket{+}_{R}\otimes \ket{+}_{R}$ & $8_c\otimes 8_c$ & Bosonic \\
		(R$_+$, R$_-$) & $\ket{+}_{R}\otimes \ket{-}_{R}$ & $8_c\otimes 8_s$ & Bosonic \\
		(R$_-$, R$_-$) & $\ket{-}_{R}\otimes \ket{-}_{R}$ & $8_s\otimes 8_s$ & Bosonic \\
		(NS$_+$, R$_+$) & $\xi_{i}\beta^i_{-1/2}\ket{0}_{NS}\otimes \ket{+}_{R}$ & $8_v\otimes 8_c$ & Fermionic \\
		(NS$_+$, R$_-$) & $\xi_{i}\beta^i_{-1/2}\ket{0}_{NS}\otimes \ket{-}_{R}$ & $8_v\otimes 8_s$ & Fermionic
	\end{tabular}
\end{center}
Imposing modular invariance leads to the following consistent theories 
\begin{center}
	\begin{tabular}{c c c c c}
		type IIA: \ & (NS$_+$, NS$_+$)& (R$_+$, R$_-$)&(NS$_+$, R$_-$)&(R$_+$, NS$_+$) \\
		type IIB: \ & (NS$_+$, NS$_+$)&(R$_+$, R$_+$)&(NS$_+$, R$_+$)&(R$_+$, NS$_+$) \\
		type 0A: \ & (NS$_-$, NS$_-$)&(NS$_+$, NS$_+$)&(R$_+$, R$_-$)&(R$_-$, R$_+$) \\
		type 0B: \ & (NS$_-$, NS$_-$)&(NS$_+$, NS$_+$)&(R$_+$, R$_+$)&(R$_-$, R$_-$)
	\end{tabular}.
\end{center}
In particular, type 0 theory contains a tachyon (with $m^2_0 = - {2\ov \a'}$), the 
same massless NSNS states, 
double the number of 
massless RR states 
(as compared to type II theory) 
and no spacetime fermions.

The corresponding torus partition function is given by ($\tau=\tau_1+i\tau_2$) 
\begin{equation}\la{203}
Z=\int_{{\F}_0}\frac{\dd\tau_1\dd\tau_2}{\tau_2}\Tr\left(e^{-2\pi\tau_2H+2\pi i \tau_1 P}\right)=\int_{\F_0}\frac{\dd^2\tau\, }{4\tau_2}
\, \big(\qq^{L_0}\bar \qq^{\bar L_0}\big) 
\ , \qquad \qquad \qq=\exp(2\pi i \tau) \ . 
\end{equation}
Separating the bosonic zero mode factor $(2\pi\sqrt{\alpha'})^{-8}\tau_2^{-5}$, 
the contribution of 8 bosonic oscillator modes 
is expressed in terms of the Dedekind $\eta$-function $\eta(\tau)=\qq^{1/24}\prod_{n=1}^\infty(1-\qq^n)$. Then 
\begin{equation}\la{206}
Z=\frac{1}{4(2\pi)^8\alpha'^4}\int_{F_0}\frac{\dd^2\tau\, }{4\tau_2^6}\, \abs{\eta(\tau)}^{-16}\, \II\ , \qquad 
\II= \Tr_{_{\text{ferm.}}}\big(\qq^{L_0}\bar \qq^{\bar L_0}\big)\ . 
\end{equation}
The fermionic contributions lead to factors of 
$\prod_{n=1}^\infty(1\pm q^{n+ a})$
(where $a=-\ha$ for NS and $0$ for R sectors) which can be expressed in terms of the Jacobi theta functions 
$ \vartheta_{ab}(0;\tau)$ defined as 
\begin{equation}
\vartheta_{ab}(z;\tau)=
\theta {\tiny \begin{bmatrix}	{a/ 2}\\ {b/ 2}
	\end{bmatrix}}(z;\tau)=\sum_{n\in\mathbb{Z}}e^{i\pi(n+a/2)^2\tau}e^{2\pi i(n+a/2)(z+b/2)}.\label{255} 
\end{equation}
One finds that the only modular invariant combinations of these factors are 
\begin{align}\la{208}
\II_{\rm II A,B}&=\frac{(\vartheta_{00}^4-\vartheta_{01}^4-\vartheta_{10}^4+\vartheta_{11}^4)(\bar\vartheta_{00}^4-\bar\vartheta_{01}^4-\bar\vartheta_{10}^4\mp\bar\vartheta_{11}^4)}{\abs{\eta}^8}\ , \\
\II_{\rm 0 A,B} &=\frac{\vartheta_{00}^4\bar\vartheta_{00}^4+\vartheta_{01}^4\bar\vartheta_{01}^4+\vartheta_{10}^4\bar\vartheta_{10}^4\mp\vartheta_{11}^4\bar\vartheta_{11}^4}{\abs{\eta}^8}\la{209}\ , \qquad \qquad \vartheta_{ab} \equiv \vartheta_{ab}(0;\tau) \ , 
\end{align}
corresponding to the above two type II and two type 0 theories. 
In the 
case of type II theories the Jacobi identity
$\vartheta_{00}^4=\vartheta_{01}^4+\vartheta_{10}^4 $
and $\vartheta_{11}(0;\tau)=0$ implies that $Z_{{\rm II A,B}}=0$.
In the type 0 case we cannot further simplify 
\begin{equation}\la{215}
Z_{\rm 0 A,B}=\frac{1}{4(2\pi)^8\alpha'^4}\int_{{\cal F}_0}\frac{\dd^2\tau\, }{4\tau_2^6}\ \frac{\vartheta_{00}^4\bar\vartheta_{00}^4+\vartheta_{01}^4\bar\vartheta_{01}^4+\vartheta_{10}^4\bar\vartheta_{10}^4\mp\vartheta_{11}^4\bar\vartheta_{11}^4}{\abs{\eta(\tau)}^{24}} \ . 
\end{equation}
Here there is an IR divergence at $\tau_2 \to \infty$ indicating the presence of the ground-state tachyon.


Let us now discuss the equivalent light-cone GS description. In the 
IIB case one starts with 
\begin{equation}\la{217}
\S=\frac{1}{\pi\alpha'}\int\dd^2\sigma~ \left(\partial_+x^i\partial_-x^i + iS_R^\alpha\partial_+S_R^\alpha+iS_L^{\alpha}\partial_-S_L^{\alpha}\right) \ , 
\end{equation} 
while in the 
type IIA case one has $S_L^{\alpha} \to S_L^{\dot \alpha}$ ($\a=1, ..., 8$). 
Here $S_{L,R}$ are worldsheet bosons and spacetime fermions
that satisfy periodic boundary conditions. 
The fermionic zero modes imply degeneracy of the ground-state 
corresponding to $8_c\oplus 8_v$ of $SO(8)$. Combining the left and right movers, 
the massless modes are 
$(8_{s/c}\oplus 8_v)\otimes(8_c\oplus 8_v)$
which is the same as in the NSR case after the GSO projection
and all higher order excitations arise from this single sector. 
The partition function vanishes since it involves $\vartheta_{11}$ 
\begin{equation}\la{219}
Z_{\rm II}=\frac{1}{4(2\pi)^8\alpha'^4}\int_{\F_0}\frac{\dd^2\tau\, }{4\tau_2^6}\ \frac{\vartheta_{11}^4\bar\vartheta_{11}^4}{\abs{\eta(\tau)}^{24}}=0. 
\end{equation}
The way to describe flat-space type 0 string theory 
starting from type II theory in the GS framework \cite{Dixon:1986iz} is to 
orbifold by a $2\pi$ rotation in a plane or equivalently by $(-1)^{F}$ where $F$ is the 
spacetime fermion number.\foot{A similar 
	light-cone GS description for a 
	$Z_2$ orbifold of type 0 string theory 
	viewed as $Z_4$ orbifold of type II string theory 
	was given in \ci{Klebanov:1999pw}.
	Related constructions with antiperiodic spacetime fermion sectors appeared in \cite{Rohm:1983aq,Atick:1988si}
	and also \ci{Blum:1997gw,Tseytlin:2001qb,Takayanagi:2001jj}.
}
Namely, one gets the ``untwisted" sector built out of type II states keeping only those
with $(-1)^F=1$, i.e. projecting out all spacetime fermions.
To preserve modular invariance one is to then add a 
``twisted" sector where the string closes only up to a transformation
by $(-1)^F$, i.e. one is to take the GS fermions $S_L$ and $S_R$ to be antiperiodic on the cylinder. 
Like in the NSR formulation that gives a non-zero normal ordering constant and thus 
a tachyonic ground state.

In the partition function, 
the projection onto bosonic modes introduces a term $\vartheta_{10}^4\bar\vartheta_{10}^4$ while the twisted sector adds $\vartheta_{00}^4\bar\vartheta_{00}^4+\vartheta_{01}^4\bar\vartheta_{01}$ and we end up with the same partition function \rf{215} as in the NSR description. 

\section{Melvin twist on pp-wave background}\label{B}

Starting with type IIB pp-wave background \rf{301} we may compactify the 
coordinate $y={u+v \ov \sqrt{2} }$ on a circle of radius $R$ and add 
a Melvin twist \rf{221} in the $(x_1, x_2)$ plane as 
\begin{equation}\la{A01} - dt^2 + dy^2 + 
\dd x_1^2+ \dd x_2^2\ \ \to\ \ - dt^2 + dy^2 + (\dd x +iqx\, \dd y) (\dd \bar{x}-iq\bar{x}\, \dd y) \ , \ \ \ \ \ \ \ x= x_1 + i x_2 \ .
\end{equation}
For compact $y$ we may fix light-cone gauge as 
\begin{equation}\la{A03}
u=\alpha'p^u\tau+wR\sigma \ .
\end{equation}
The transverse string fluctuations now acquire mass 
(cf. \rf{302})
\begin{equation}\la{A04}
\mu^2=f^2\big[\alpha'^2(p^u)^2-(wR)^2\big]\,.
\end{equation} 
The derivation of the spectrum is similar to the case of a 
Melvin twist in flat space in section \rf{Melvin}. 
The zero-mode momentum in $y$ direction is now $p^y= { m \ov R}$ and thus we get 
(below $\xi= qR$ and $\gamma= w \xi$ as in \rf{222},\rf{225}, cf. \rf{226}) 
\begin{align}
&\hat p^u=\tfrac{\hat \mm }{R} + \sqrt{\left(\tfrac{\hat \mm }{R}\right)^2+\left(\tfrac{w R}{\alpha'}\right)^2+\tfrac{2}{\alpha'}\hat{N}}\ , \qquad \qquad \hat \mm \equiv \mm-\xi(\hat{J}_b+\ha \hat{J}_f)
\la{A12}\ , \\ 
& \hat H=- \hat p^v=-\ha \tfrac{\mm}{R}+ \tfrac{1}{2}\sqrt{\left(\tfrac{\hat \mm }{R}\right)^2+\left(\tfrac{w R}{\alpha'}\right)^2+\tfrac{2}{\alpha'}\hat{N} } \ , \la{A13} \\
&\hat E^2=(\hat p^t)^2=\left(\tfrac{\hat \mm }{R}\right)^2+\left(\tfrac{w R}{\alpha'}\right)^2+\tfrac{2}{\alpha'}\hat{N}\ . \la{A14}
\end{align}
Here the expressions for $\hat N$ (before normal ordering) and the bosonic contribution to the 
angular momentum operator $\hat{J}_b$ are given by
\begin{align}\la{A15}
\hat{N}&=\hat{N}_x+\hat{N}_\perp+\hat{N}_f\ , \\
\hat{N}_x&=\sum_{n=1}^{\infty}\big( \omega_{n-\gamma}\alpha_{n+}^\dagger \alpha_{n+}+\omega_{n+\gamma}\alpha_{n-}\alpha_{n-}^\dagger+\omega_{n-\gamma}\tilde{\alpha}_{n-}\tilde{\alpha}_{n-}^\dagger 
+\omega_{n+\gamma}\tilde{\alpha}_{n+}^\dagger \tilde{\alpha}_{n+}\big) 
+ \omega_{\gamma}(\alpha_{0-}\alpha^\dagger_{0-} + \tilde{\alpha}_{0+}^\dagger \tilde{\alpha}_{0+})
\no \\ 
\hat{N}_\perp&=\tfrac{1}{2}\sum_{i=3}^8\Big[\sum_{n=1}^{\infty}\omega_{n}(\alpha_{n}^{i\dagger} \alpha_{n}^{i}+ \alpha_{n}^{i}\alpha_{n}^{i\dagger}+\tilde{\alpha}_n^{i\dagger}\tilde{\alpha}_n^i+\tilde{\alpha}_n^i\tilde{\alpha}_n^{i\dagger})+\tfrac{1}{2}\mu (\alpha_0^{i\dagger}\alpha_0^i+\alpha_0^i\alpha_0^{i\dagger}+\tilde{\alpha}_0^{i\dagger}\tilde{\alpha}_0^i+\tilde{\alpha}_0^i\tilde{\alpha}_0^{i\dagger})\Big]
 \no\\ 
\hat{J}_b&=\sum_{n=1}^\infty\big (\alpha_{n+}^\dagger \alpha_{n+} - \alpha_{n-}^\dagger \alpha_{n-}+\tilde{\alpha}_{n+}^\dagger \tilde{\alpha}_{n+}- \tilde{\alpha}_{n-}^\dagger \tilde{\alpha}_{n-}\big)+\tilde{\alpha}_0^\dagger\tilde{\alpha}_0-\alpha_0^\dagger \alpha_0\ , \la{A18}
\end{align}
where frequencies are given by $\omega_n =\sqrt{n^2 + \mu^2}$ as in \rf{308} and the modes are again labelled by their sign in $\hat{J}_b$. 
The fermionic part of the GS Lagrangian in the light-cone gauge takes the form (we follow \cite{Metsaev:2002re} and use the same 
spinor notation, i.e. $\Pi=\gamma^1\bar{\gamma}^2\gamma^3\bar{\gamma}^4$, etc.)
\begin{equation}\la{A23}
\mathcal{L}=i \theta^1\bar{\gamma}^v(\partial_+-\tfrac{1}{2}q\gamma^{12}\partial_+y)\theta^1+i \theta^2\bar{\gamma}^v(\partial_--\tfrac{1}{2}q\gamma^{12}\partial_-y)\theta^2 -2\mu\theta^1\bar{\gamma}^v\Pi\theta^2+\dots.
\end{equation}
Introducing the commuting projectors
$
P^\pm =\frac{1\mp i\gamma^{12}}{2} , \ \ \PP^\pm =\frac{1\pm \Pi}{2} $
we may define 
\begin{equation}\la{A26}
\Lambda^{1,2}=(e^{\frac{i}{2} qy} P^+ + e^{-\frac{i}{2} qy}P^-)\theta^{1,2}\ .
\end{equation}
and thus rewrite \rf{A23} as 
\begin{equation}\la{A27}
\mathcal{L}= i \Lambda^1\bar{\gamma}^v\partial_+\Lambda^1+i \Lambda^2\bar{\gamma}^v\partial_-\Lambda^2 -2\mu\Lambda^1\bar{\gamma}^v\Pi\Lambda^2+\dots \ . 
\end{equation}
Then we can use the expressions in \cite{Metsaev:2002re} accounting for the shifted periodicities arising from the phase factors. 
The fermionic equations of motion split under $ \PP^\pm$ into \begin{equation}\la{A28}
\partial_+\Lambda^1\mp \mu \Lambda^2=0, \qquad\qquad \partial_-\Lambda^2\pm \mu \Lambda^1=0\ , 
\end{equation}
and finally the fermionic contributions to the oscillator number operator \rf{A15} and the angular momentum in the (1,2)-plane which appears in 
\rf{A12} are found to be 
\begin{align}\no 
&\hat{N}_f=\sum_{n=1}^\infty\big( \omega_{n-\frac{\gamma}{2}}\beta_{n+}^\dagger \beta_{n+}+\omega_{n+\frac{\gamma}{2}}\beta_{n-}^\dagger \beta_{n-}+\omega_{n+\frac{\gamma}{2}}\tilde{\beta}_{n+}^\dagger \tilde{\beta}_{n+}+\omega_{n-\frac{\gamma}{2}}\tilde{\beta}_{n-}^\dagger \tilde{\beta}_{n-}\big) +\omega_{\frac{\gamma}{2}}(\beta_0^\dagger \beta_0+\tilde{\beta}_0^\dagger\tilde{\beta}_0)\\\la{A30}
&\hat{J}_f
=\sum_{n=1}^\infty(\beta_{n+}^\dagger \beta_{n+}-\beta_{n-}^\dagger \beta_{n-}+\tilde{\beta}_{n+}^\dagger \tilde{\beta}_{n+}-\tilde{\beta}_{n-}^\dagger\tilde{\beta}_{n-})+\tilde{\beta}_{0}^\dagger\tilde{\beta}_{0}-\beta_{0}^\dagger \beta_{0} \ , 
\end{align}
where the 
tilde distinguishes between left- and right-moving modes and the subscript 
$\pm$ corresponds to the eigenvalues of $\gamma^{12}$. 
As the fermions couple to $y$ with ``charge'' $\ha q$, as in \rf{226}, 
there is an extra $\ha$ coefficient between $\hat J_b$ and $\hat J_f$ 
in $\hat \mm$ \rf{A12}. 
Setting $\xi=1$ and taking the limit $R\to 0$ like in the case of the flat space Melvin twist 
in section \ref{Melvin}, and then 
applying T-duality\foot{Without applying T-duality we get the spectrum of type 0A theory in the background T-dual to the 
	pp-wave \rf{301}, i.e. in a generalisation of the fundamental string (or F-model \ci{Horowitz:1994ei}) type background supported also by a combination of RR fluxes.}
we end up with the spectrum of type 0B theory in the pp-wave background \rf{301}, 
i.e. with the sum of the untwisted and twisted sectors discussed in section \ref{8D}.

\section{Type 0B effective action expanded near pp-wave background
}\label{C}

In the case of type IIB theory \cite{Metsaev:2002re} one can reproduce the low-energy part of the 
string spectrum in pp-wave background \rf{301} by expanding the string theory effective action or the corresponding equations 
for the target space fields. Here we shall do the same in the type 0B case. 

The effective action of type 0B string theory 
can be reconstructed from the string scattering amplitudes \cite{Klebanov:1998yya} 
\footnote{Additional information about the structure of the action 
	can be inferred indirectly by imposing consistency with T-duality \cite{Meessen:2001wk}.
	Here we use the sign conventions of \cite{Becker:2007zj} and the 
	normalisation of \cite{Klebanov:1998yya}. In \cite{Klebanov:1998yya} the action is defined with an additional minus-sign in front of the action and $\kappa=\ha$. To connect to \cite{Meessen:2001wk} we have to introduce an overall minus sign in front of the metric and another one in front of the Ricci scalar, set $\kappa={1 \ov\sqrt{2}}$ and rescale $T\to \sqrt{2}T$ and $F_n\to \sqrt{2}F_n$.}
\begin{align}\la{B01}
&S=\tfrac{1}{2\kappa^2}\Big\{\int\dd^{10}x\sqrt{-G}\Big( e^{-2\Phi}\Big[R+4\partial_\mu\Phi\partial^\mu\Phi-\tfrac{1}{4}(\partial_\mu T\partial^\mu T + m^2_0 T^2)-\tfrac{1}{2}\abs{H_3}^2\Big] \\
& - \tfrac{1}{4}e^{h(T)}(\abs{F_1^+}^2+|{\tilde{F}_3^+}|^2+|{\tilde{F}_5}|^2)- \tfrac{1}{4}e^{-h(T)} (\abs{F_1^-}^2+|{\tilde{F}_3^-}|^2)\Big)+\ha \int F_5\wedge F_3^-\wedge B_2\Big\} + ...\ .\no
\end{align}
Compared to type IIB theory where we have a graviton $g$, a dilaton $\Phi$, a Kalb-Ramond field $B$ with field strength $H$ and RR potentials $C_0$, $C_2$ and $C_4$ with field strengths $F_1$, $F_3$ and $F_5$ (the later restricted by a self-duality constraint), 
here we have an additional tachyon 
scalar $T$ with $m^2_0 = - {2\ov \a'} $ and a doubled set $F^{\pm}_1$, $F^{\pm}_3$, $F_5$ 
of the massless RR fields\foot{Here there is one copy of $F_5$ but it is not self-dual. We use the notation $|F_p|^2 = {1 \ov p!} F_{\mu_1 ...\mu_p} F^{\mu_1 ...\mu_p}$.} which appear in the combinations 
\begin{equation}\la{B02}
\tilde{F}_n=F_n - H_3 \wedge C_{n-3}\ , \ \ \ \qquad n=3,5\ . 
\end{equation}
The coupling of the tachyon to the RR fields is given by an odd function 
\be\la{c3} h(T)=T+\order{T^3} \ . \ee
One should note that the effective action including the 
tachyon 
is not, strictly speaking, well defined as a 
derivative expansion. In particular, 
the tachyon effective action reconstructed from on-shell amplitudes and expanded in powers of derivatives 
is ambiguous \cite{Banks:1991sg,Tseytlin:1991bu}: one can use field redefinitions and the 
leading-order equation 
$\a' \Box T = -2 T $, and thus cannot distinguish between $(-{\a'\ov 2} \Box)^n T$ and $T $ factors
in the action.

Our aim will be to compute the spectrum of small fluctuations 
of the tachyon and massless fields in \rf{B01} near 
the pp-wave background \rf{301}. Since the latter has only metric and (self-dual) 
$F_5$ being non-trivial, 
the cubic interaction terms in \rf{B01} involving the RR fields which are relevant for the study of quadratic fluctuations are
\begin{equation}\la{B05}
\begin{split}
S_3 =\tfrac{1}{4\kappa^2}\int \Big[ C_2^+\wedge H_3 \wedge \star F_5 +F_5\wedge F_3^- \wedge B_2 -TF_5\wedge\star F_5\Big]\ . 
\end{split}
\end{equation}
The relevant terms in the resulting (linearised) equations of motion may be written as 
\begin{align}\la{B06}
&R_{\mu\nu}=\tfrac{1}{4\cdot4!} ( 1 + T) \Big(F_{\mu\alpha\beta\gamma\delta}F_{\nu}^{\alpha\beta\gamma\delta}-\tfrac{1}{10}G_{\mu\nu}F_{\alpha\beta\gamma\delta\epsilon}F^{\alpha\beta\gamma\delta\epsilon}\Big)\ , \\\la{B07}
&\na^2\Phi=0 \ , \qquad 
(\na^2 -m_0^2)T=4f(F_{v1234}+F_{v5678}) \ , \qquad 
\na^2 C^\pm = 0 \ , \\
&\na^\mu F_{\mu\nu\rho}^\pm =\mp 4f H^{\alpha\beta\gamma}(\delta^{u1234}_{\nu\rho\alpha\beta\gamma}+\delta^{u5678}_{\nu\rho\alpha\beta\gamma})\ , \\\la{B12}
&\na^\mu H_{\mu\nu\rho}=2fF^{\alpha \beta\gamma+}(\delta^{u1234}_{\nu\rho\alpha\beta\gamma}+\delta^{u5678}_{\nu\rho\alpha\beta\gamma})-2fF^{\alpha\beta\gamma-}(\delta^{u1234}_{\nu\rho\alpha\beta\gamma}+\delta^{u5678}_{\nu\rho\alpha\beta\gamma})\ , \\\la{B14}
&\na^\mu F_{\mu\nu\rho\sigma\eta}=-4f\partial^\alpha T(\delta^{u1234}_{\nu\rho\sigma\eta\alpha}+\delta^{u5678}_{\nu\rho\sigma\eta\alpha})\ .
\end{align}
Here $\delta^{\m_1...\m_n}_{\n_1 ...\n_n}= \delta^{\m_1}_{[\n_1}... \delta^{\m_n}_{\n_n]}$. 
Introducing $F_3=\frac{1}{\sqrt 2} (F_3^+-F_3^-)$ and $\bar{F}_3=\frac{1}{\sqrt 2}(F_3^++F_3^-)$ we can write 
the equations for $F_3^\pm, H$ as:
\begin{align}\la{B15}
&\na^\mu \bar{F}_{\mu\nu\rho}=0\ , \qquad 
\na^\mu F_{\mu\nu\rho}=-2^{5/2}f H^{\alpha\beta\gamma}(\delta^{u1234}_{\nu\rho\alpha\beta\gamma}+\delta^{u5678}_{\nu\rho\alpha\beta\gamma})\ , \\
&\na^\mu H_{\mu\nu\rho}=2^{3/2}fF^{\alpha\beta\gamma}(\delta^{u1234}_{\nu\rho\alpha\beta\gamma}+\delta^{u5678}_{\nu\rho\alpha\beta\gamma})\ . 
\end{align}
For the dilaton $\Phi$ and the RR scalars $C^\pm$, we get the massless wave equation (cf. \rf{325})
\begin{equation}\la{B18}
\na^2 \Phi=\frac{1}{\sqrt{-G}}\partial_\mu(\sqrt{-G}G^{\mu\nu}\partial_\nu\Phi)=\left(2\partial_u\partial_v+f^2x^2_i\partial_v^2+\partial_i^2\right)\Phi=0 \ , 
\end{equation}
which is solved as in \cite{Metsaev:2002re} or in \rf{325}--\rf{326} with the effective 
light-cone Hamiltonian $H=-p_v= f ( \sum_{i=1}^8 a^{\dagger }_i a_i+4)$ 
with the lowest energy state having
$H_0= f\mathcal E_0, \ \mathcal E_0=4$. 
This matches 
what one finds in the type 0B spectrum: the 
contribution $4$ arises from either 
4 fermionic excitations of the untwisted sector (for $\Phi, C^+$ as in type IIB case \cite{Metsaev:2002re}) 
or from the first fermionic excitation in the twisted 
sector once combined with the normal ordering constant \rf{319} expanded in $\a'$.

Following the notation in \cite{Metsaev:2002re} 
for a field satisfying a similar wave equation with an extra linear $\del_v$ term one can read off the lowest light-cone energy value $H_0$ as 
\begin{equation}\la{B23}
(\na^2 +2if\, c\, \partial_v)\phi=0 \qquad \rightarrow \qquad{H}_0=f\mathcal{E}_0 \ , \qquad \mathcal{E}_0 = 4+c\ .
\end{equation}
Fixing the gauge $\bar{C}_{v\mu}=0$ for the field $\bar{F}_3$ 
one finds as in \cite{Metsaev:2002re} that 
$\na^2 \bar{C}_{ij}=0$ and thus 28 massless modes with $\mathcal{E}_0=4$. In the following we need to distinguish two sets of indices $i={1,2,3,4}$ and $i'={5,6,7,8}$ as these couple to the 5-form flux independently. 
For $(F_3,H_3)$ one finds that $\na^2 C_{ii'}=\na^2 B_{ii'}=0$
while modes with indices from the same set as e.g. 
$A_{12}=C_{12}+\sqrt{2}iB_{34}$ and $\bar{A}_{12}=C_{12}-\sqrt{2}iB_{34}$ satisfy 
\begin{equation}\la{B28}
(\na^2 -4if\partial_v)A_{12}=0, \quad (\na^2 +4if\partial_v)\bar{A}_{12}=0 \ \ \to \ \ \mathcal{E}_0=4\pm2 \ . 
\end{equation}
The most complicated sector is that of the fluctuations of the 
metric, the 5-form and the tachyon. Since $T$ is coupled to $F_5$ in \rf{B01}, the non-zero $F_5$-background 
leads to mixing of the fluctuations of $F_5$ with the tachyon. 
Expanding to linear order in fluctuations 
\begin{align}\no
G_{\mu\nu}\to G_{\mu\nu}+h_{\mu\nu},\quad C_{\mu\nu\rho\sigma}\to C_{\mu\nu\rho\sigma}+c_{\mu\nu\rho\sigma}\ , \quad 
R_{\mu\nu}\to R_{\mu\nu}+r_{\mu\nu},\quad F_{\mu\nu\rho\sigma\eta}\to F_{\mu\nu\rho\sigma\eta}+a_{\mu\nu\rho\sigma\eta}, 
\end{align}
and choosing the light-cone gauge
$ h_{v\mu}=0,\ \ c_{v\mu\nu\rho}=0 $, 
the linearised Einstein equation takes the form
\begin{align}
r_{\mu\nu}=&fa_{\mu}^{\alpha\beta\gamma\delta}\delta_{\nu\alpha\beta\gamma\delta}^{u1234}+fa_{\nu}^{\alpha\beta\gamma\delta}\delta_{\mu\alpha\beta\gamma\delta}^{u1234}-f^2 h^{\alpha\alpha'}G^{\beta\beta'}G^{\gamma\gamma'}G^{\delta\delta'}\delta_{\mu\alpha\beta\gamma\delta}^{u1234}\delta_{\nu\alpha'\beta'\gamma'\delta'}^{u1234} + (1234\to 5678)\no \\
&+8f^2T\delta_{\mu\nu}^{uu}-G_{\mu\nu}f(a_{v1234}+a_{v5678})\ , \la{B32} \\
r_{\mu\nu}=&\tfrac{1}{2}\left(-\na^2 h_{\mu\nu}+\na_{\mu}\na^\rho h_{\rho\nu}+\na_{\nu}\na^\rho h_{\rho\mu}-\na_{\mu}\na_{\nu}h^\rho_\rho+2R_{\mu\rho\sigma\nu}h^{\rho\sigma}+R_{\mu\rho}h^{\rho}_\nu+R_{\nu\rho}h^{\rho}_\mu\right),\no
\end{align}
with the non-zero components being (background curvature is $R_{iuui}=-f^2, \ \ R_{uu}=8f^2$) 
\begin{align}
r_{uu}&=\te 2f (a_{u1234}+a_{u5678})-2f^2\sum_{i=1}^8h^{ii}+8f^2T+f^2x_I^2(a_{v1234}+a_{v5678})\no \\r_{ij}&=f(a_{v1234}-a_{v5678})\delta_{ij}, \quad r_{i'j'}= f(a_{v5678}-a_{v1234})\delta_{i'j'}, \quad r_{ui}=f(\epsilon_{ijkl} a_{vujkl}+a_{i5678}),\no \\ r_{ui'}&=f(\epsilon_{i'j'k'l'}a_{vuj'k'l'}+a_{i'1234}),\quad r_{ii'}=f(\epsilon_{ijkl}a_{vi'jkl}+\epsilon_{i'j'k'l'}a_{vij'k'l'})\ .
\end{align}
where $\epsilon_{1234}=\epsilon_{5678}=1$ and repeated indices $i$ or $i'$ are summed over. 
All other components vanish.
As in \cite{Metsaev:2002re} the ($vv$) component of the Einstein equation vanishes 
and thus gives the zero-trace condition for the transverse modes
$h_i^i+h_{i'}^{i'}=0$.
\footnote{This also follows directly from the 10d Weyl invariance of the 
	action for $F_5$ and thus tracelessness of its stress tensor in \rf{B06}.} 
The ($vi$) components vanish 
as well, which determines the non-dynamical modes $h_{ui}$ via 
$\partial^uh_{ui}+\partial^jh_{ij}+\partial^{j'}h_{ij'}=0$. 
The ($vu$) component implicitly fixes $h_{uu}$ via 
$\partial^uh_{uu}+\partial^ih_{iu}+\partial^{i'}h_{i'u}=0$. 
The off-diagonal 
($ij$) and ($i'j'$) components give the free field equations for $h_{ij}$, $h_{i'j'}$
\begin{equation}\la{B40}
\na^2 h_{ij}=0, \quad \quad \na^2 h_{i'j'}=0\ \quad \rightarrow \quad \mathcal{E}_0=4 \ .
\end{equation}
The remaining metric fluctuations are mixed with $C_4$ potential fluctuations via 
\begin{align}\la{B41}
\na^2 h_{ij} &= -2f\partial_v(c_{1234}-c_{5678})\delta_{ij}\ , \qquad 
\na^2 h_{i'j'} = -2f\partial_v(c_{5678}-c_{1234})\delta_{i'j'}\ , \no \\
\na^2 h_{ii'} &=-2f\partial_v(\epsilon_{ijkl}c_{i'jkl}+\epsilon_{i'j'k'l'}c_{ij'k'l'}) \ . 
\end{align} 
The equation for $F_5$ expanded to first order is 
\begin{align}\no
\na^\mu a_{\mu\nu\rho\sigma\eta}-G^{\alpha\beta}\na^{\mu}h_{\nu\alpha}\, F_{\mu\beta\rho\sigma\eta}-&G^{\alpha\beta}\na^{\mu}h_{\rho\alpha}\, F_{\mu\nu\beta\sigma\eta}-G^{\alpha\beta}\na^{\mu} h_{\sigma\alpha}\, F_{\mu\nu\rho\beta\eta}-G^{\alpha\beta}\na^{\mu}h_{\eta\alpha}\, F_{\mu\nu\rho\sigma\beta}\\
=&-2f\partial^\alpha T(\delta^{u1234}_{\nu\rho\sigma\eta\alpha}+\delta^{u5678}_{\nu\rho\sigma\eta\alpha}) \ . 
\end{align}
Here we can gauge fix $c_{v\mu\nu\rho}=0$ and then eliminate $c_{u\mu\nu\rho}$. Then
$
\na^2 c_{iji'j'}=0 \rightarrow \mathcal{E}_0=4$
while the remaining physical fluctuations satisfy 
\begin{align}
\na^2 c_{1234}&=4f\partial_v(h_{11}+h_{22}+h_{33}+h_{44})-4f\partial_vT \ ,\no \\ \na^2 c_{5678}&=4f\partial_v(h_{55}+h_{66}+h_{77}+h_{88})-4f\partial_vT\ , \la{B46} \\
\na^2 c_{i'jkl}&=4f\epsilon_{ijkl}\partial_vh_{ii'}\ , \qquad 
\na^2 c_{ij'k'l'}=4f\epsilon_{i'j'k'l'}\partial_vh_{ii'}\ .\no 
\end{align} 
To simplify these equations let us introduce the following combinations of fields (using $h^i_i=0$) 
\begin{equation}\la{B47}
\begin{split}
h&=h_{11}+h_{22}+h_{33}+h_{44}\ , \qquad\quad\ h'=h_{55}+h_{66}+h_{77}+h_{88}=- h \ , \\
g_a&=h_{11}+h_{aa}-h_{bb}-h_{cc}\ , \qquad \ \ \ \ (a,b,c)=\sigma(2,3,4)\ ,\\
g_{a'}&=h_{55}+h_{a'a'}-h_{b'b'}-h_{c'c'}\ , \qquad (a',b',c')=\sigma(6,7,8)\ ,\\
c_{ii'} &= \epsilon_{ijkl}c_{i'jkl}+\epsilon_{i'j'k'l'}c_{ij'k'l'}\ , \qquad \ \ \ \bar{c}_{ii'} = \epsilon_{ijkl}c_{i'jkl}-\epsilon_{i'j'k'l'}c_{ij'k'l'}\ . 
\end{split} 
\end{equation}
Then $\bar{c}_{ii'}$, $g_a$ and $g_{a'}$ satisfy free massless equations 
as in \rf{B40}
and thus yield 22 modes with 
$\mathcal{E}_0=4$. 
The field $c_{ii'}$ combines with $h_{ii'}$ and yields 16 modes of both $\mathcal{E}_0=2$ and $\mathcal{E}_0=6$.
It remains to solve the equations for the 4 interdependent 
fluctuations 
$c=c_{1234}$, $c'=c_{5678}$, $h$ and $T$.
They satisfy the equations
\begin{align}\la{B48}
&\na^2 c = 4f\partial_v h-4f\partial_vT\ , \qquad 
\na^2 c' = -4f\partial_v h -4f\partial_vT\ \ , \no \\
&\na^2 h = -8f\partial_v(c-c') \ ,\ \ \qquad (\na^2-m_0^2) T = 4f\partial_v (c+c') \ .
\end{align}
These equations can be obtained also by 
directly expanding the action \rf{B01} to quadratic order in fluctuations 
\begin{align}\la{B49}
S[{h,h',c,c',T}]=\tfrac{1}{2\kappa^2}\int \dd ^{10}x \sqrt{-G}\Big[&\tfrac{1}{16}h\na^2 h+\tfrac{1}{16}h'\na^2 h'+\tfrac{1}{4}T(\na^2-m_0^2)T+\tfrac{1}{4}c\na^2 c+\tfrac{1}{4}c'\na^2 c'\no \\
&-2fT\partial_v(c+c')+f(h-h')\partial_v(c-c')\Big]\ , 
\end{align}
and imposing the zero-trace condition $h=-h'$ after the variation. 
Introducing 
\be 
c_-= c-c' \ , \qquad \qquad c_+=c+c' \ , \la{b49}
\ee
we find that $c_-$ and $h$ satisfy
\be 
\na^2 c_- = 8f\partial_v h \ , \qquad \na^2 h = -8f\partial_v c_- \ , \ \ \ \ \qquad
\na^2 (h\pm i c_-) = \pm 8if \del_v ( h\pm i c_-) \ , 
\ee
and thus using \rf{B23} $h \pm i c_- $ have $\mathcal{E}_0=0$ and $8$.
From \rf{B48} the 
remaining fields $c_+$ and $T$ satisfy a coupled system 
\be 
\na^2 c_+ = -8f\partial_v T \ , \qquad (\na^2-m^2_0) T = 4f\partial_v c_+ \ , 
\ee
implying $ (\na^2-m^2_0)\na^2 T = - 32 f^2 \del_v^2 T$. It can be diagonalised by introducing the new fields 
\iffa 
\begin{equation}\la{B50}
\begin{pmatrix} 
\na^2 & 2^{5/2}i \mub \\
-2^{5/2}i \mub & \na^2 -m_0^2
\end{pmatrix}
\begin{pmatrix}
c_+ \\
\sqrt{2}T
\end{pmatrix}
=0 \ , \qquad \qquad \mub\equiv p^u f \, , \ \ \ \ m^2_0 = - \te {2\ov \a'} \ . 
\end{equation} 
\fi 
\begin{align}
&K^\pm=4f \del_v T+ \hat {\cal M }_\mp \, c_+ \ , \qquad \hat { \cal M}_\pm \equiv \ha {m_0^2}\pm \sqrt{\tfrac{1}{4}{m^4_0} - 32f^2\del_v^2}\ , 
\la{B52} \\ 
&(\na^2-\hat { \cal M}_\pm )K^\pm=0 \ \ \quad \rightarrow \quad\ \ \mathcal{E}_0 
= 4+ {m^2_0 \ov 4 fp^u } \pm \sqrt{\frac {m_0^4}{16(fp^u)^2}+8} \ , 
\la{b53}
\end{align}
where in \rf{b53} we used the momentum representation and $p^u=p_v$. 

The resulting spectrum may be summarised as follows:
{\small 
	\begin{center}
		\begin{tabular}{|c|c|c|c|}
			\hline
			Fields & Components & $ \mathcal{E}_0 $ & \# \\
			\hline \hline
			$\Phi$ & $\Phi$ & 4 & 1 \\
			\hline
			$F_1^+$, $F_1^-$ & $C^+$, $C^-$ & 4 & 2 \\
			\hline 
			
			\multirow{4}{4em}{\begin{center}$F_3^+$, $F_3^-$, $H_3$\end{center}} & $\bar{C}_{ij} $,$\bar{C}_{i'j'} $,$\bar{C}_{ii'} $ & $4$ & 28 \\ 
			& $C_{ii'}$, $B_{ii'} $ & 4 & 32 \\
			& $A_{ij}$, $A_{i'j'}$ & $2$ & 12 \\
			& $\bar{A}_{ij}$, $\bar{A}_{i'j'}$ & $6$ & 12 \\
			\hline
			\multirow{9}{4em}{\begin{center}$G$, $F_5$, $T$\end{center}} & $h_{ij}$, $h_{i'j'}$, $g_{a}$, $g_{a'}$ & 4 & 18 \\
			& $c_{iji'j'}$, $\bar{c}_{ii'}$ & 4 & 52 \\
			& $h_{ii'}+ic_{ii'}$ & 2 & 16 \\
			&$h_{ii'}-ic_{ii'}$ & 6 & 16 \\
			& $h+ic_-$ & 0 & 1 \\
			& $h-ic_-$ & 8 & 1 \\
			& $K^+$ & $ 4+ {m^2_0 \ov 4 fp^u } + \sqrt{\frac {m_0^4}{16(fp^u)^2}+8} 
			$ & 1 \\ 
			& $K^-$ & $4+ {m^2_0 \ov 4 fp^u } - \sqrt{\frac {m_0^4}{16(fp^u)^2}+8} 
			$ & 1 \\
			\hline
		\end{tabular}
\end{center} }
Here we gave the values of the rescaled light-cone energy $ \mathcal{E}_0$, with the eigenvalue of the light-cone Hamiltonian being 
$H=-p^v = f \mathcal{E}_0$. The effective mass or dispersion relation for 
the corresponding state may be written as in \rf{305} or \rf{366},\rf{345}, i.e. ($E=p^t$, $p^u=p_v= {p^y+E\ov \sqrt 2}$)
\be
m^2 = E^2-(p^y)^2= 2 fp^u \mathcal{E}_0 \ . \la{c29}
\ee
The spectrum of the subset of fields (dilaton $\Phi$, metric $G$, $H_3$, one copy of $F_1$ and $F_3$ and the 
self-dual part of $F_5$) present also in type IIB theory, i.e.
of the untwisted sector of type 0B theory, is of course the same as in \cite{Metsaev:2002re}. 

The remaining states -- the 63 extra 
``massless'' ($\mathcal{E}_0=4$) modes and ``massive'' mixtures $K^\pm$ of the tachyon and the RR 5-form should belong to the twisted sector of type 0 states. 
For the ``lower'' or ground-state mode $K^+$ we get from \rf{c29}
\begin{align}
m^2 = &8 fp^u + \ha {m^2_0 } + \ha m_0^2 \sqrt{1 + 128 m_0^{-4} ( f p^u)^2 } \no\\
=&-\tfrac{2}{ \a'} + 8 fp^u - 16 \a'\left(f p^{u}\right)^2+128 \a'^3(fp^{u})^4 + ... \ . \la{c299}
\end{align}
We used that $m_0^2=- {2\ov \a'}$ and expanded in $\a'$ or in powers of ${f p^u\ov m_0^2} = - \ha \mu $
where $\mu= \a' p^u f$ is the parameter used in the exact string spectrum (cf. \rf{308},\rf{312}). 
Comparing \rf{c299} to $m^2 = - m_0^2 A_8 (\mu)$ in \rf{344} we observe that the 
first two terms match while the coefficient of the leading order 
correction $\a' (f p^u)^2$ is - 16 in \rf{c299} and $-16 \ln 2$ in \rf{344}. 
This mismatch is not too surprising as here the starting point was the low-energy effective action \rf{B01} 
 which (as was mentioned below \rf{c3}) contains an ambiguity
 when expanded in derivatives of the tachyon 
 field\foot{For example, the 
 $T F_5 F_5$ coupling can be replaced by 
 $ -\ha \a' \del^2 T F_5 F_5$, etc.}
 and also does not include higher $\a'$ corrections. 
 
 Similarly, for $K^-$ we get 
\be
m^2 = 8 fp^u + \ha {m^2_0 } - \ha m_0^2 \sqrt{1 + 128 m_0^{-4} ( f p^u)^2 } 
= 8 fp^u + 16 \a' \left(f p^{u}\right)^2 - 128 \a'^3 (fp^{u})^4 + ... \ . \la{c288}
\ee
Thus to leading order $K^-$ is effectively massless and should 
combine with the other $63$ modes of $\mathcal{E}_0=4$ in the above table 
 to complete the first excited level of the twisted sector. 
 Indeed, the states on the first excited level in the twisted sector are created by applying a pair of fermionic creation 
 operators in \rf{307},\rf{309} to the vacuum and thus the corresponding analog of the ground-state mass relation \rf{344} is 
 \be \la{c300}
 m^2= \tfrac{2}{ \a'} \big[ A_8(\mu) + \sqrt{ 1 + \mu^2}\, \big]= 8 fp^u + \OO\big(\a' (fp^u)^2\big) \ , \ee
 i.e. has the same leading term as in \rf{c288}. 
 Here we expanded in powers of $\mu= \a' p^u f$ and the flat-space tachyon part $-1$ in $A_8$ cancelled against 
 the leading excited state contribution.
 The mismatch of subleading order $\a'$ terms is again attributed to 
 the fact that the action \rf{B01} does not contain $\a'$ corrections.
 One may, in fact, turn this around and try to use the information about the exact pp-wave spectrum 
 to fix the structure of higher $\a'$ terms in the effective action.


\

\bibliographystyle{JHEP}
\small\baselineskip 12pt
\bibliography{Type.bib}

\end{document}